\documentclass[%
 aip,
 amsmath,amssymb,
 reprint,%
]{revtex4-1}

\usepackage{graphicx}
\usepackage{dcolumn}
\usepackage{bm}
\usepackage{notes2bib}
\usepackage[utf8]{inputenc}
\usepackage[T1]{fontenc}
\usepackage{etoolbox}
\usepackage{amsmath}
\usepackage{amssymb}
\usepackage{braket}
\usepackage{color}
\usepackage{ulem}
\usepackage{mathbbol}
\usepackage{cancel}
\usepackage{graphicx}
\definecolor{darkgreen}{RGB}{40,110,5}

\usepackage{ulem}  
\definecolor{orange}{rgb}{1,0.5,0}
\definecolor{darkgreen}{RGB}{40,110,5}

\newcommand{\oE}[1]{{\color{white}\sout{#1}}}
\newcommand{\bfepsilon}{\mathbf{\epsilon}\hspace{-4.5pt}\mathbf{\epsilon}}
\newcommand{\bfmu}{\mathbf{\mu}\hspace{-5.1pt}\mathbf{\mu}}

\usepackage{mathalfa}
\DeclareMathAlphabet{\mathdutchcal}{U}{dutchcal}{m}{n}

\makeatletter
\def\@email#1#2{%
 \endgroup
 \patchcmd{\titleblock@produce}
  {\frontmatter@RRAPformat}
  {\frontmatter@RRAPformat{\produce@RRAP{*#1\href{mailto:#2}{#2}}}\frontmatter@RRAPformat}
  {}{}
}%
\makeatother
\begin{document}

\preprint{AIP/123-QED}

\title{Formally exact fluorescence spectroscopy simulations for mesoscale molecular aggregates with $\mathbf{N^0}$ scaling
}
\author{Tarun Gera}
\affiliation{Department of Chemistry, University of Texas at Austin, Austin, Texas 78712, USA}
\author{Alexia Hartzell}
\affiliation{Department of Chemistry, University of Texas at Austin, Austin, Texas 78712, USA}
\author{Lipeng Chen}
\affiliation{Zhejiang Laboratory, Hangzhou 311100, China}
\author{Alexander Eisfeld}
\affiliation{Max Planck Institute for the Physics of Complex Systems, N\"{o}thnitzer Str 38, Dresden, Germany}
\affiliation{
Institute of Theoretical Physics, TUD Dresden University of Technology, 01062, Dresden, Germany
}
\author{Doran I. G. B. Raccah}%
\affiliation{Department of Chemistry, University of Texas at Austin, Austin, Texas 78712, USA}
\email{doran.raccah@utexas.edu}

\begin{abstract}

We present a size-invariant (i.e., $N^0$) scaling algorithm for simulating fluorescence spectroscopy in large molecular aggregates. We combine the dyadic adaptive hierarchy of pure states (DadHOPS) equation-of-motion with an operator decomposition scheme and an efficient Monte Carlo sampling algorithm to enable a formally exact, local description of the fluorescence spectrum in large molecular aggregates. Furthermore, we demonstrate that the ensemble average inverse participation ratio (IPR) of DadHOPS wave functions reproduces the delocalization extent extracted from fluorescence spectroscopy of J-aggregates with strong vibronic transitions. This work provides a computationally efficient framework for fluorescence simulations, offering a new tool for understanding the optical properties of mesoscale molecular systems.
\end{abstract}

\maketitle

\section{Introduction}
Absorption and fluorescence spectroscopy probe the excited-state dynamics of molecular materials and are sensitive to both electronic energy levels and electron-vibrational coupling. Because of the spectral congestion in the absorption and fluorescence spectra of most molecular materials, theoretical simulations and modeling play an important role in unveiling the underlying physical mechanisms that control the excited-state dynamics reported by the spectral line shapes and peak positions. Absorption spectra can often be reproduced by using high-temperature, mixed quantum classical, short-time, or perturbative approximations (or a combination thereof).\cite{Dutta2025, Kapral2006, Jansen2024, Fetherolf2017} Fluorescence spectroscopy, however, is a more challenging observable: the entanglement between the electronic and vibrational degrees of freedom that occurs during the relaxation towards thermal equilibrium in the excited-state causes electronic localization (or a decrease in the coherence length) that is essential to reproducing the dynamic Stokes shift in many molecular materials.\cite{ Dutta2025,ma2015forster} As a result, many of the approximate methods that perform well for absorption spectra do not reproduce fluorescence spectra. 

Formally exact methods, which account for the interaction between the electronic and vibrational degrees of freedom, face significant challenges in simulating fluorescence for mesoscale molecular aggregates composed of thousands of molecules or more. Techniques such as multiconfiguration time-dependent Hartree (MCTDH),\cite{Meyer1990,Beck2000,Burghardt2003,Richings2015} multilayer MCTDH (ML-MCTDH),\cite{Wang2003,Wang2015,Manthe2008,Vendrell2011}  multi-configuration Ehrenfest,\cite{Shalashilin2009,Shalashilin2010,Chen2019} and ab initio multiple spawning\cite{Ben1998,Ben2000} propagate the combined electronic and vibrational wave function for the whole molecular aggregate. Extending these techniques to even moderately sized molecular aggregates often requires additional approximations: for example, assuming each molecule's vibrations can be characterized by a single harmonic oscillator.\cite{Martinazzo2006} Alternatively, formally exact solutions to an open quantum system Hamiltonian, such as hierarchical equations of motion (HEOM),\cite{Tanimura2006,tanimura2020numerically} time-dependent density matrix renormalization group theory (TD-DMRG),\cite{Marston2002,White2004} and quasi-adiabatic path integrals (QUAPI),\cite{Makri1992,makri1995} simplify dynamics by evolving the reduced density matrix of electronic states coupled to an effective thermal environment, parameterized to reproduce the characteristic  frequencies and relaxation timescales of molecular vibrations. Recently, techniques like modular path integrals\cite{makriCommunicationModularPath2018,makriModularPathIntegral2018} and tensor contraction\cite{borrelliExpandingRangeHierarchical2021,yanEfficientPropagationHierarchical2021,somozaDissipationAssistedMatrixProduct2019,tamascelliEfficientSimulationFiniteTemperature2019,strathearnEfficientNonMarkovianQuantum2018b} have demonstrated improved scaling for formally exact solutions to open quantum systems.

Non-Markovian quantum state diffusion (NMQSD)\cite{Diosi1997,Strunz1998PRA} unravels the reduced density matrix into an ensemble of wave functions that can be independently time-evolved. By propagating wave functions instead of the density matrix, NMQSD reduces memory requirements but necessitates generating numerous realizations to construct the ensemble. The hierarchy of pure states (HOPS)\cite{Suess2014,Hartmann2017,Gerhard2015JCP} reformulates the NMQSD equations in a numerically efficient way. 
In HOPS, the stochastic state vectors are coupled to a hierarchy of auxiliary vectors,  subject to complex-valued correlated Gaussian noise determined by the coupling to the environment.
This hierarchy is very similar to the one used in the HEOM method, which consists of coupled auxiliary density matrices.
The HEOM can be derived from HOPS in a straight-forward manner.\cite{Suess2015}
Thus, HOPS is formally exact in the same sense, and under the same assumptions, as HEOM while offering  numerical advantages due to propagating a stochastic wave function.

The \textit{dyadic} HOPS\cite{Chen2022, Chen2022JCP} is a reformulation of HOPS, particularly suitable for perturbation theory of external interactions.
By propagating the bra and ket sides of the reduced density matrix separately via the HOPS equation, dyadic HOPS establishes a direct connection to the non-linear response function formalism. However, like other exact methods, dyadic HOPS encounters computational challenges when applied to large molecular aggregates. The recent development of adaptive HOPS (adHOPS) \cite{Varvelo2021, Varvelo2023, Citty2024} addresses this limitation by leveraging the locality of physical wave functions, achieving size-invariant ($N^0$) scaling  for sufficiently large aggregates. 
Building on these principles, Dyadic adaptive HOPS (DadHOPS) was introduced,\cite{Gera2023} incorporating the adaptive framework into the dyadic formulation. Ref.~\onlinecite{Gera2023} leveraged Monte Carlo sampling over an initial state decomposition of the dipole autocorrelation function calculated with DadHOPS to achieve size-invariant ($N^0$) scaling linear absorption simulations for mesoscale molecular aggregates. 

In this paper, we present a method for simulating fluorescence spectroscopy in large molecular aggregates that achieves size-invariance. We demonstrate that the initial state decomposition method (renamed as Excitation Operator Decomposition in this work) can be applied simultaneously to both bra and ket states, allowing the optical response functions of a material to be reconstructed by Monte Carlo sampling spatially local contributions. The resulting DadHOPS calculations for fluorescence spectroscopy have size-invariant scaling for sufficiently large aggregates. Finally, we demonstrate that the average delocalization length (or coherence number) of the electronic excited states calculated from the participation ratio of the HOPS wave function reproduces that extracted from fluorescence measurements, which further supports the value of analyzing HOPS wave function ensembles to understand the physical mechanism underlying the observed dynamics of the reduced density matrix.

\section{Theoretical Background}

\subsection{Hamiltonian} \label{Sec:Hamiltonian}

We use an open quantum system approach, where the total matter Hamiltonian is composed of the system Hamiltonian $(\hat{H}_{\mathrm{S}})$, the bath (or environment) Hamiltonian $(\hat{H}_{\mathrm{B}})$ and the Hamiltonian that describes the interaction $(\hat{H}_{\mathrm{SB}})$ between them: 
\begin{equation}\label{eq:Htot}
\hat{H}_\mathrm{M}=\hat{H}_{\mathrm{S}}+\hat{H}_{\mathrm{B}}+\hat{H}_{\mathrm{SB}}.
\end{equation}

The system studied in this work consists of a molecular aggregate containing $N$ interacting pigments. 
The $n^{\textrm{th}}$ pigment is described as  a two level system  with  ground ($\ket{g_n}$) and excited ($\ket{e_n}$) electronic states and corresponding energy levels $E_n^g$ and $E_n^e$. In the following discussion, we restrict ourselves to the single-exciton manifold, where the Hamiltonian can be expressed as
\begin{equation}
\label{eq:H_sys}
\hat{H}_{\mathrm{S}}=E_g|g\rangle\langle{g}|+\sum_{n=1}^{N}E_n|n\rangle\langle{n}|+{\sum_{n=1}^N}\sum_{m\neq n}^{N}J_{nm}|n\rangle\langle{m}|
\end{equation}
with the common ground state $\ket{g}=\prod_n \ket{g_n}$ at energy $E_g=\sum_n E_n^g$ and singly excited state $\ket{n}=\ket{e_n}\prod_{m\neq n}\ket{g_m}$ at energy $E_n=E_n^e+\sum_{m\neq n}E_m^g$. Each pigment has an independent thermal reservoir to account for interactions with intra- and intermolecular vibrations. We call these degrees of freedom a bath and the corresponding bath Hamiltonian 
\begin{equation}
    \hat{H}_{\mathrm{B}}=\sum_{n=1}^N\sum_{q_n} \hbar\omega_{q_n}\hat{b}_{q_n}^{\dagger}\hat{b}_{q_n}
\end{equation}
consists of an infinite set of harmonic oscillators where $\hat{b}_{q_n}^{\dagger}$ and $\hat{b}_{q_n}$ are the creation and annihilation operators for the $q^{\textrm{th}}$ mode of the $n^{\textrm{th}}$ pigment with frequency $\omega_{q_n}$.  \\ 
The linear coupling between electronic state and bath is accounted for by the system-bath Hamiltonian
\begin{equation}
\hat{H}_{\mathrm{SB}} = \sum_{n=1}^N \hat{L}_n \sum_{q_n}\Lambda_{q_n}\left(\hat{b}_{q_n}^{\dagger}+\hat{b}_{q_n}\right) 
\end{equation}
where $\hat{L}_n= \ket{n}\bra{n}$ is a system-bath coupling operator and $\Lambda_{q_n}$ is the exciton–bath coupling strength of the $q^{\textrm{th}}$ mode of the $n^{\textrm{th}}$ pigment. Introducing the bath spectral density $\kappa_n(\omega) =\pi \sum_{q_n}|\Lambda_{q_n}|^2\delta(\omega-\omega_{q_n})$, the bath correlation function is given by
\begin{equation}
\begin{aligned}
\alpha_n(\tau)=\frac{1}{\pi}\int_0^{\infty}\!\!\!\mathrm{d}\omega\, \kappa_n(\omega)\Big( \coth\big(\frac{\beta \hbar\omega}{2}\big) \cos (\omega \tau) -i \sin(\omega \tau)\Big)
\label{eq:bath_corr}
\end{aligned}
\end{equation}
at the inverse temperature $\beta=\frac{1}{\textrm{k}_B T}$.

In this study, we focus on fluorescence spectroscopy in an open quantum system and neglect internal conversion, as it typically occurs on a much longer timescale (nanoseconds) than those considered here (picoseconds). 

Fluorescence spectroscopy involves two types of light-matter interactions: the first is the incoming field responsible for exciting the material, which is treated classically, while the second is the spontaneously emitted light, which is treated quantum mechanically. The corresponding total Hamiltonian, incorporating both the material and radiation degrees of freedom, is given by
\begin{eqnarray}
    \hat{H}_{\textrm{tot}}(t)&=&\hat{H}_\mathrm{M} +\hat{H}_\mathrm{F}+\hat{H}_\mathrm{L}(t)\nonumber\\
    &=&\hat{H}_0 + \hat{H}_\mathrm{L}(t)
\end{eqnarray}
where $\hat{H}_\mathrm{F}$ is the Hamiltonian for the free field and $\hat{H}_\mathrm{L}(t)$ is the light-matter interaction Hamiltonian. In the following we consider the emission into a mode with frequency $\omega_R$ and use
\begin{equation}
    \hat{H}_\mathrm{F} = \hbar\omega_R\left(\hat{a}_R^{\dagger}\hat{a}_R + \frac{1}{2}\right)\label{Eq:Hamiltonian_field}
\end{equation}
where the field is emitted in mode $R$ with $\hat{a}^{\dagger}_R$ and $\hat{a}_R$ as the corresponding creation and annihilation operators.
Finally, the Hamiltonian for the interaction of light with matter consists of two terms, 
\begin{equation}
    \hat{H}_\mathrm{L}(t)=-\hat{\bfmu}\cdot\mathbf{E}_{I}(t) -\hat{\bfmu}\cdot\hat{\mathbf{E}}_{R}.\label{Eq:Hamiltonian_LM}
\end{equation}
The first term accounts for the interaction between matter and the incoming classical electric field
\begin{equation}
  \mathbf{E}_I(t)=E_I(t)\bfepsilon_I e^{i\omega_I t}  +E_I^*(t)\bfepsilon^*_I e^{-i\omega_I t}
\end{equation}
with amplitude $E_I(t)$, direction of polarization $\bfepsilon_I$, and frequency $\omega_I$. 
The spontaneously emitted radiation mode
\begin{equation}
  \hat{\mathbf{E}}_R
  =E_R\bfepsilon_R\, \hat{a}_R 
  +E_R^* \bfepsilon_R^*\, \hat{a}^{\dagger}_R
\end{equation}
is treated quantum mechanically with its initial state in vacuum and with amplitude $E_R=-i\left(2\pi\hbar\omega_R/\Omega\right)^{1/2}$ for frequency $\omega_R$ and box normalization volume $\Omega$.
The collective electronic transition dipole moment operator  is given by
\begin{equation}
\hat{\bfmu}=\hat{\bfmu}^+ + \hat{\bfmu}^-.
\end{equation}
For fluorescence, we confine the collective transition dipole moment operator to the subspace consisting of the ground state and singly excited states, yielding

\begin{equation}
\label{eq:collective_Transition_Operator}
 \hat{\bfmu}^+=\sum_{n=1}^N \bfmu_n \,|n\rangle\langle{g}|\quad \mathrm{and}\quad \hat{\bfmu}^-=\sum_{n=1}^N \bfmu_n^* \,|g\rangle\langle{n}|
\end{equation}
where $\bfmu_n= \bra{g_n} \hat{\bfmu}_n\ket{e_n}$ is the dipole transition vector of the $n^\mathrm{th}$ molecule and $\hat{\bfmu}^+=(\hat{\bfmu}^-)^\dagger$.
For convenience, we introduce scalar transition operators
 \begin{equation}
 \label{eq:defV+-}
     \hat{V}^{+}_{j}=\frac{(\hat{\bfmu}^{+}\cdot\bfepsilon_{j})}{D_{j}}\quad \mathrm{and}\quad \hat{V}^{-}_{j}=(\hat{V}^{+}_{j})^\dagger,\quad (j= R,I)
 \end{equation}
  which are normalized by the collective transition weights
 \begin{equation}
 \label{eq:transition_weights}
     D_{j}=\sqrt{\sum_{n=1}^N|\bfmu_n\cdot\bfepsilon_{j}|^2}.
 \end{equation}

 The light-matter Hamiltonian defined in Eq.~\eqref{Eq:Hamiltonian_LM} follows the standard treatment of spontaneous emission in non-linear spectroscopy.\cite{mukamel1995principles,Mukamel1985} A treatment of the quantized radiation field (Eq.~\eqref{Eq:Hamiltonian_field}) capable of describing higher-order processes (e.g., reabsorption) would include a sum over an infinite set of modes across all frequencies. As we will see below, however, the photo-emission process described at third-order involves the trace over the number operator of a specific radiation mode at frequency $\omega_R$ and because only two interactions are allowed with the quantized field Hamiltonian, that radiation mode is the only one to contribute to the emission rate. As a result, we can, without approximation, reduce the quantized field to only that radiation mode.
 
 \subsection{Correlation function for Fluorescence }
Optically excited fluorescence is a process of spontaneous light emission that depends linearly on the incoming light intensity. Despite having a linear dependence on the weak incoming field intensity, fluorescence calculations require a third order response function. In order to calculate the fluorescence response of a material, we start with the initial total density matrix $(\hat{\rho}_{\mathrm{tot}}(0))$ at equilibrium
\begin{equation}
\hat{\rho}_{\mathrm{tot}}(0)=\hat{\rho}_\mathrm{S}^{\textrm{eq}}\otimes\hat{\rho}_\mathrm{B}^{\textrm{eq}}\otimes\hat{\rho}_R^{\textrm{vac}}
\end{equation}
where the system is in its ground state, $\hat{\rho}_\mathrm{S}^{\textrm{eq}}=\ket{g}\bra{g}$, the bath is in a thermal equilibrium state $\hat{\rho}_\mathrm{B}^{\textrm{eq}}=  e^{-\beta{\hat{H}}_{\mathrm{B}}}/\mathrm{Tr}_{\mathrm{B}}\left\lbrace e^{-\beta{\hat{H}}_{\mathrm{B}}}\right\rbrace$, and the emitted radiation mode is in its vacuum state $\hat{\rho}_R^{\textrm{vac}}=\ket{0_R}\bra{0_R}$.

\begin{figure}
\includegraphics[width=1\linewidth]{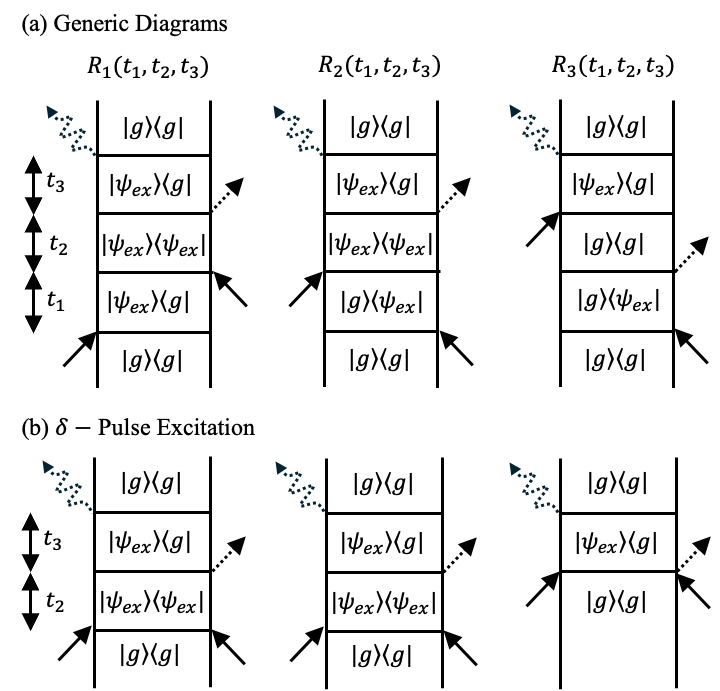}
\caption{Double-sided Feynman diagrams for pathways contributing to spontaneous light emission. (a) Generic diagrams. (b) Diagrams for the special case of a delta function excitation. The solid arrows represent interactions with the incoming light while the dotted arrows denote interactions with the emitted mode.  }\label{Fig:FD}
\end{figure}

The rate of photon emission into mode $\hat{\mathbf{E}}_R$ after two interactions with the classical light field is approximately given by (for details see Appendix \ref{App:photon_emission_rate} and the derivation of Eq.~(9.10) in Ref.~\onlinecite{mukamel1995principles})  
\begin{align}\label{Eq:Approx_Pt}
    P(t)&=\frac{1}{\hbar^4}\frac{2\pi\hbar\omega_R}{\Omega}S_{\textrm{SLE}}(\omega_I,\omega_R,t).
\end{align}
The most relevant part of this expression

\begin{multline}
S_{\mathrm{SLE}}(\omega_I,\omega_R,t) = 2 D_I^2D_R^2 \operatorname{Re} 
\int_0^{\infty} \int_0^{\infty} \int_0^{\infty} dt_1 \, dt_2 \, dt_3 \\
\Big( E_I(t-t_1-t_2-t_3)E^*_I(t-t_2-t_3) e^{-i\omega_I t_1 - i\omega_R t_3} R_1(t_1,t_2,t_3) \\
+ E^*_I(t-t_1-t_2-t_3)E_I(t-t_2-t_3) e^{i\omega_I t_1 - i\omega_R t_3} R_2(t_1,t_2,t_3) \\
+ E^*_I(t-t_1-t_2-t_3)E_I(t-t_3) 
      e^{i\omega_I t_1 - i(\omega_I-\omega_R)t_2 - i\omega_R t_3} \\
\quad \times R_3(t_1,t_2,t_3) \Big)
\label{Eq:Response_function_SLE}
\end{multline}

contains the collective transition weights defined in Eq.~(\ref{eq:transition_weights})
and the response functions
\begin{align}
    R_1(t_1, t_2, t_3) &= \left< \hat{V}^{-}_{I}(t_1) \hat{V}^{+}_{R}(t_1+t_2) \right. \nonumber\\
    & \quad \times \hat{V}^{-}_{R}(t_1+t_2+t_3)\hat{V}^{+}_{I}(0)\hat{\rho}(0)\big>\nonumber\\
    R_2(t_1, t_2, t_3) &= \left< \hat{V}^{-}_{I}(0)\hat{V}^{+}_{R}(t_1+t_2) \right. \nonumber\\
    & \quad \times \hat{V}^{-}_{R}(t_1+t_2+t_3)\hat{V}^{+}_{I}(t_1)\hat{\rho}(0)\big>\nonumber\\
    R_3(t_1, t_2, t_3) &= \left<\hat{V}^{-}_{I}(0)\hat{V}^{+}_{R}(t_1) \right. \nonumber\\
    & \quad \times \hat{V}^{-}_{R}(t_1+t_2+t_3)\hat{V}^{+}_{I}(t_1+t_2)\hat{\rho}(0)\big> \label{Eq.4ptCorrFn}
\end{align}
with $\hat{\rho}(0)=\hat{\rho}_\mathrm{S}^{\textrm{eq}}\otimes\hat{\rho}_\mathrm{B}^{\textrm{eq}}$ and 
\begin{equation}
     \hat{V}^{\pm}_{j}(t)=\hat{U}_\mathrm{M}^{\dagger}(t)\hat{V}^{\pm}_{j}\hat{U}_\mathrm{M}(t)
 \end{equation}
 where
 \begin{equation}
     \hat{U}_\mathrm{M}(t)=e^{-iH_\mathrm{M}t/\hbar}
 \end{equation}
 is the time-evolution with respect to the material Hamiltonian. 
  The double-sided Feynman diagrams for these three pathways are drawn in Fig.~\ref{Fig:FD}(a). 
 
To simplify our calculations below, we make two assumptions: (1) the incoming light source is a short temporal pulse we can approximate by a delta function $E_I(t)=E_I \delta(t)$, and (2) the measurement time of fluorescence is longer than the optical dephasing time of the material. Under the delta function approximation, the two excitation operations must occur simultaneously for each pathway (see Fig.~\ref{Fig:FD}(b)). Furthermore, since in the $R_3(t_1,t_2,t_3)$ pathway the excitation operations are the first and third interactions, the corresponding amplitude under the delta-function approximation (assumption (1)) decays on the timescale of the optical dephasing time and can be neglected (assumption (2)). The other two terms in Eq.~\eqref{Eq:Response_function_SLE} become identical (as can also be seen from the double-sided Feynman diagrams in Fig.~\ref{Fig:FD}(b)), which simplifies the expression to 
\begin{align}
    S_{\textrm{SLE}}&(\omega_I,\omega_R,t)
    \equiv \mathdutchcal{F}(\omega_R,t)\\
    &=4D_I^2D_R^2|E_I|^2\,\textrm{Re}\int_0^{\infty}dt_3 e^{-i\omega_Rt_3} R_1(0,t-t_3,t_3)\label{Eq:S_SLE_R1_R2}
\end{align}
which no longer depends on the excitation frequency. 
Because  the response function evolves more slowly during the population time ($t_2$) than during the coherence time ($t_3$), when the measurement time ($t=t_2+t_3$) is long compared to the optical dephasing time, we can approximate $R_1(0,t-t_3,t_3)\approx R_1(0,t_2,t_3) $ which further simplifies Eq.~\eqref{Eq:S_SLE_R1_R2} to
\begin{equation}
    \mathdutchcal{F}(\omega_R,t_2) = 4D_I^2D_R^2|E_I|^2\,\textrm{Re}\left[\int_0^\infty R(t_2,t_3) \mathrm{e}^{-i \omega_R t_3} dt_3\right]\label{Eq:FourierTransformR}
\end{equation}
where we have used the short-hand notation 
\begin{equation}
\label{eq:def_R(t2,t3)}
\begin{split}
  R(t_2,t_3)\equiv & R_1(0,t_2,t_3) \\
  =&\big<  
 \hat{V}^{-}_{R}(t_2+t_3)\hat{V}^{+}_{I}\hat{\rho}(0)\hat{V}^{-}_{I}\hat{V}^{+}_{R}(t_2) \big>.
  \end{split}
\end{equation}

Naively, Eq.~\eqref{Eq:FourierTransformR} can be interpreted as the fluorescence spectrum of the photons emitted at the time $t_2$. Care should be taken, however, since  $\mathdutchcal{F}(\omega_R,t_2)$ is not formally an observable: the fluorescence observed at the measurement time $t$ should be integrated over the range of possible population times ($t_2=t-t_3$). This subtlety does not impact the results or the method presented below and we neglect it in the calculations that follow.  All results presented in the following sections are computed for a fixed waiting time (\(t_2\)). Nevertheless, one can simulate the time evolution of the fluorescence signal by adjusting the waiting time of the calculation; an example of such a result is shown in Appendix~\ref{App:TDF}.
\oE{[fine]}

\subsection{Dyadic HOPS for Fluorescence Spectroscopy}\label{Sec:DadHOPS}

Here, we use the dyadic Hierarchy of Pure States (HOPS) method\cite{Chen2022JCP} to calculate the non-linear response function describing fluorescence spectroscopy.
 For completeness, we briefly recapitulate the original derivation\cite{Chen2022JCP} in Appendix~\ref{AppI} specialized to the case of the $3^{rd}$-order response function for fluorescence spectroscopy.

 Within the dyadic HOPS approach the response function of Eq.~(\ref{Eq.4ptCorrFn}) can be obtained using  trajectories $\ket{\tilde{\psi}(t_2,t_3;\mathbf{z^*})}$ in a doubled (dyadic) \textit{system} Hilbert space that depends on stochastic processes, indicated by the argument $\mathbf{z^*}$:
\begin{equation}
    R(t_2,t_3)=\mathcal{M}_\mathbf{z}\Big\{
    I(t_2;\mathbf{z^*})\,\bra{\tilde{\psi}(t_2,t_3;\mathbf{z^*})}\tilde{F}\ket{\tilde{\psi}(t_2,t_3;\mathbf{z^*})}\Big\}\label{Eq:Generic_Dyadic_R}
\end{equation}
where  $\mathcal{M}_\mathbf{z}\{\cdots\}$ denotes the average over trajectories and 
 $\tilde{F}=\begin{pmatrix}
    0 & 0\\
    \hat{F} & 0
\end{pmatrix}$  with $\hat{F}=\hat{V}^-_R$ being the final light matter interaction operator appearing in $R_1$ of Eq.~(\ref{Eq.4ptCorrFn}).
The wave function trajectories are given by
\begin{equation}
    \ket{\tilde{\psi}(t_2,t_3;\mathbf{z^*})}= \tilde{G}(t_3;\mathbf{z^*})\tilde{V}_3\tilde{G}(t_2;\mathbf{z^*})\tilde{V}_2\tilde{V}_1\ket{\tilde{\psi}_0}\label{eq:dyadic_wf}
\end{equation}
where the dyadic initial state,
\begin{eqnarray}  
\ket{\tilde{\phi}_0}=\begin{pmatrix}\ket{g}\\ \ket{g} \end{pmatrix},  
\end{eqnarray}  
is normalized prior to propagation
\begin{eqnarray}  
  \ket{\tilde{\psi}_0} =\frac{1}{\sqrt{2}}\ket{\tilde{\phi}_0}= \frac{1}{\sqrt{2}}\begin{pmatrix}\ket{g}\\ \ket{g} \end{pmatrix},  
\end{eqnarray} 
the dyadic light matter interaction operators are given by
\begin{equation}
\begin{array}{ccc}
\label{eq:V1,V2,V3}
\tilde{V}_1 = \begin{pmatrix}
        \hat{V}_I^+ & 0\\
        0 & \mathbb{I}
        \end{pmatrix}, &
\tilde{V}_2 = \begin{pmatrix}
         \mathbb{I} & 0\\
        0 & \hat{V}_I^+
        \end{pmatrix}, &
\tilde{V}_3 = \begin{pmatrix}
         \mathbb{I} & 0\\
        0 & \hat{V}_R^-
        \end{pmatrix},
\end{array}
\end{equation}
and the (stochastic) propagators $\tilde{G}(t;\mathbf{z^*})$ imply that the state on the right is propagated using the formally exact dyadic HOPS equation.\cite{Chen2022JCP}  

The normalization factor $I(t_2;\mathbf{z^*})$, accounts for the change in magnitude of the response function arising from the light-matter interactions when the time-evolution is performed using normalized wave functions. To propagate a normalized wave function, we normalize the wave function after each action of the light-matter interaction operators in Eq.~\eqref{eq:dyadic_wf}. In Eq.~\eqref{Eq:Generic_Dyadic_R}, to compensate for these normalizations, we  multiply each trajectory by 
\begin{align}
I(t_2;\mathbf{z^*})
&=
||\tilde{V}_3 \ket{\tilde{\psi}^{(0)}(t_2;\mathbf{z^*})}||^2
\cdot ||\tilde{V}_2 \ket{\tilde{\psi}_1}||^2 \nonumber\\
&\times||\tilde{V}_1 \ket{\tilde{\psi}_0}||^2\cdot||\ket{\tilde{\phi}_0}||^2
\end{align}
 where $||\cdot||^2$ denotes the squared  L2 norm, i.e., $||\ket{\tilde{\psi}}||^2 = \braket{\tilde{\psi}|\tilde{\psi}}$, and $\ket{\tilde{\psi}_1} = \tilde{V}_1 \ket{\tilde{\psi}_0} / |\tilde{V}_1 \ket{\tilde{\psi}_0}|$,
and $\ket{\tilde{\psi}^{(0)}(t_2; \mathbf{z}^*)}$ is obtained from the normalized HOPS equation with initial state 
 \begin{equation}
 \begin{aligned}
     \ket{\tilde{\psi}^{(0)}(0;\mathbf{z^*})} = \ket{\tilde{\psi}_2}&=\tilde{V}_2 \ket{\tilde{\psi}_1}/ ||\tilde{V}_2 \ket{\tilde{\psi}_1}||\\
     &= \tilde{V}_2\tilde{V}_1\ket{\tilde{\psi}_0}/||\tilde{V}_2 \tilde{V}_1 \ket{\tilde{\psi}_0}||.
 \end{aligned}
  \end{equation}
We propagate the wave function using the \textit{normalized} non-linear HOPS equation 
\begin{flalign}
\begin{aligned}
\label{eq:NormNonLinearHops}
\hslash \partial_t&\vert \tilde{\psi}^{(\Vec{k})}_t \rangle 
=  \big(-i\tilde{H}_\mathrm{S} - \Vec{k} \cdot \Vec{\gamma} -\Gamma_t \big)\ket{\tilde{\psi}^{(\Vec{k})}_t}\\  
+& \sum_{n} \tilde{L}_{n} (z^*_{n,t}+ \xi_{n,t}) \ket{\tilde{\psi}^{(\Vec{k})}_t}   \\ 
+ &\sum_{n}^{N}\sum_{j_n}^{N_{\textrm{mode}}} k_{n,j_n} \gamma_{n,j_n} \tilde{L}_{n}  \vert \tilde{\psi}^{(\Vec{k} -\Vec{e}_{n, j_n})}_t \rangle \\
- &\sum_{n}^{N}\sum_{j_n}^{N_{\textrm{mode}}} \left(\frac{g_{n,j_n}}{\gamma_{n,j_n}}\right)(\tilde{L}^{\dagger}_{n} - \langle\tilde{L}^{\dagger}_{n}\rangle_{t}) \vert \tilde{\psi}^{(\Vec{k}+\Vec{e}_{n, j_n})}_t\rangle
\end{aligned}
\end{flalign}
where we indicate the time dependence by a lower index $t$ and we do not explicitly write the dependence on $\mathbf{z^*}$ of the wave function.
In the above equation we have 
$
\begin{array}{ccc}
\tilde{H}_\mathrm{S}=\begin{pmatrix}
        \hat{H}_\mathrm{S} & 0\\
        0 & \hat{H}_\mathrm{S} \\
        \end{pmatrix} & , & \tilde{L}_n=\begin{pmatrix}
        \hat{L}_n & 0\\
        0 & \hat{L}_n \\
        \end{pmatrix}
\end{array}
$, noise dependent expectation values 
\begin{equation}
    \langle\tilde{L}^{\dagger}_{n}\rangle_{t} = \langle \tilde{\psi}^{(\vec{0})}(t;\mathbf{z^*}) \vert \tilde{L}^{\dagger}_{n}\vert \tilde{\psi}^{(\vec{0})}(t;\mathbf{z^*}) \rangle
\end{equation}
drift in the noise 
\begin{equation}
    \xi_{t,n} = \frac{1}{\hbar}\int_{0}^{t} ds \alpha^{*}_{n}(t-s) \braket{\tilde{L}^{\dagger}_{n}}_s
\end{equation}
and stochastic normalization correction factor 
\begin{flalign}
\begin{aligned}
\label{eq:normcorr}
    \Gamma_t = &\sum_{n} \braket{\tilde{L}_{n}}_{t} \textrm{Re}[z^*_{n,t}+ \sum_{j_n}\xi_{j_n,t}] \\
    - &\sum_{n,j_n}  \textrm{Re}\Big[\Big(\frac{g_{n,j_n}}{\gamma_{n,j_n}}\Big)\braket{\tilde{\psi}^{(\Vec{0})}(t)|\tilde{\psi}^{(\Vec{k})}(t)}\Big] \\
    + &\sum_{n,j_n}  \braket{\tilde{L}^{\dagger}_n}_{t} \textrm{Re}\Big[\Big(\frac{g_{n,j_n}}{\gamma_{n,j_n}}\Big)\braket{\tilde{\psi}^{(\Vec{0})}(t) | \tilde{\psi}^{(\Vec{k})}(t) }\Big].
\end{aligned}
\end{flalign} 
 In deriving the above equation the bath correlation function Eq.~(\ref{eq:bath_corr}) is approximated as a sum of exponential modes
 \begin{equation}
\label{eq:alpha_dl_coarse}
    \alpha_n(t) = \sum_{j_n=1}^{N_{\textrm{mode}}} g_{n,j_n} e^{-\gamma_{n,j_n} t/\hbar} 
\end{equation}
with complex $g_{n,j_n}$ and $\gamma_{n,j_n}$. 
 The exponents $\gamma_{n,j_n}$ for each bath correlation function mode for all $N$ pigments are elements of vector $\Vec{\gamma}$. 
 
 The physical wave function $\vert \tilde{\psi}^{(\vec{0})}(t) \rangle$ represents the state of the system degrees of freedom; all the other values in index $\Vec{k}$ correspond to auxiliary wave functions that represent the non-Markovian bath interactions and the total basis is a direct product ($\mathbb{A} \bigotimes \mathbb{S}$) of the set of auxiliary wave functions ($\mathbb{A}$) and the set of pigment states ($\mathbb{S}$). 
 The coupled set of equations Eq.~(\ref{eq:NormNonLinearHops}) has to be suitably truncated;\cite{Suess2014,ZhBeEi18_134103_} in this case we apply the triangular truncation criterion ($\{\Vec{k} \in \mathbb{A}:  \sum_{n,j_n} k_{j_n} \leq k_{\textrm{max}}\}$) to constrain the hierarchy to the converged finite depth denoted as $k_{\textrm{max}}$. In the dyadic HOPS equations, the light-matter interaction operators ($\tilde{V}_j$) always act on \textit{all} 
 $\ket{\tilde{\psi}^{(\Vec{k})}(t)}$.

In the calculations below, we make use the following simplifications: (i) nearest-neighbor electronic coupling \( J_{nm} = J_{n,n\pm1} \) in Eq.~\eqref{eq:H_sys}, where \( m = n \pm 1 \); (ii) unit transition dipole moments that are parallel and aligned with the electric field polarization; and (iii) we define the specific value of $t_2$ and then scan over $t_3$ values from 0 to $t_3^\mathrm{max}$  in steps of the propagation time step $\Delta t$. We apply zero-padding to the response function $R(t_2, t_3)$ at the end, where the total length of the zero-padded function is given by $(\Delta t\cdot\omega_\mathrm{res})^{-1}$, and $\omega_\mathrm{res}$ determines the frequency resolution. In Section \ref{sec:Application}, to mitigate noise in the calculated spectra arising from the combination of zero-padding and the incomplete cancellation of the response function at long times, the aggregate spectra are smoothed by applying a cosine apodization window to the time-domain response function,
\begin{equation}
    \zeta(t) = 
    \begin{cases}
      \cos(\frac{\pi}{2}t/t_3^{\mathrm{max}}) &  t\leq t_3^{\mathrm{max}} \\
      0        & t>t_3^{\mathrm{max}}
    \end{cases}
\end{equation}
which reduces to zero at the final time point of the computed response function ($t_3^{\mathrm{max}}$).\cite{hoch1996nmr} 

\section{Size-invariant Fluorescence simulation}
 Fluorescence can be simulated using dyadic HOPS, but the computational cost increases drastically with system size. 
 In this section, we demonstrate how to achieve size-invariant scaling for fluorescence calculations: first, we reconstruct the total response function from spatially localized contributions and then employ a Monte Carlo sampling to efficiently capture the spatially localized contributions.
 
For simplicity, in this section we describe the system-bath coupling using a Drude-Lorentz spectral density: 
\begin{equation}
W_n(\omega) = \frac{2 \lambda \gamma \omega}{\omega^2  + \gamma^2}\label{Eq:DL_spectral}
\end{equation}
with real-valued $\lambda$ and $\gamma$
and employ the high-temperature approximation to reduce  the corresponding bath correlation function to
\begin{equation}
    \alpha_n(t) = (2 \lambda / \beta - i \lambda \gamma) e^{-\gamma t/\hbar}. \label{Eq:DL_bath}
\end{equation}
 The parameters used in these calculations are $\lambda = 35 $ cm$^{-1}$, $\gamma = 50$ cm$^{-1}$, and $T=295$ K. The method described here is not limited to simple spectral densities, and a more realistic bath correlation function is used in Section \ref{sec:Application}.

 \subsection{Constructing local response functions}\label{Sec:EOD}
The optical response functions of a material can be 
reconstructed by summing over spatially local contributions for each double-sided Feynman diagram, using an Excitation Operator Decomposition (EOD) method. The approach we present here is the site basis analog to previous work using secular Redfield theory\cite{Redfield1957} where the material response functions were decomposed as a sum over pathways where each excitation operator excites a single eigenstate of the system Hamiltonian.\cite{Yang2002,Ishizaki2009}

Let us first note that the operators $V^+_I$ and  $V^-_I$ , describing the interaction of the classical field on the bra and ket state,  can be decomposed as a sum over clusters of molecules ($\mathbf{d}$). Here, $\mathbf{d}$ denotes a collection of molecular indices such that the union of all sets $\mathbf{d}$ is the set of all molecules $\{1,\dots,N\}$. Examples of possible decompositions include the single-site decomposition $\mathcal{D}_\mathrm{single}=\{ \{1\}, \dots, \{N\} \}$ where each molecule is individually excited and the N-site decomposition $\mathcal{D}_\mathrm{all}=\{\{ 1, \dots, N\} \}$ where all molecules are collectively excited (i.e., there is no decomposition). Starting from Eq.~(\ref{eq:defV+-}), we can write 
\begin{align}
    V^{\pm}_I = \sum_{n=1}^N \hat{\sigma}^{\pm}_n
\quad
\mathrm{with}
\quad
    \hat{\sigma}^+_n = \ \frac{\bfmu_n \cdot\bfepsilon_{I}}{D_I} \,\vert n\rangle\langle g \vert \label{Eq:EOD_first_step}
\end{align}
and $\hat{\sigma}^-_n=(\hat{\sigma}^+_n)^\dagger$.
Because of linearity of the summation one can also write
\begin{align}
\label{eq:decomposition}
    V^{\pm}_I = \sum_{\mathbf{d}\in \mathcal{D}} \hat{\sigma}^\pm_{\mathbf{d}}
\qquad
\mathrm{with}
\qquad
    \hat{\sigma}^\pm_{\mathbf{d}} = \sum_{n \in\mathbf{d} } \hat{\sigma}^\pm_n.
\end{align}

The total response is reconstructed by summing over local response functions obtained using the EOD for initial excitations of the ket and bra states. With the definitions Eq.~(\ref{eq:decomposition}), the response function Eq.~(\ref{eq:def_R(t2,t3)}) can be written as 
\begin{equation}
\label{eq:ISD_R_alex}
    R(t_2,t_3)=\sum_{\mathbf{d}_\mathrm{K}\in \mathcal{D}_\mathrm{K}} 
    \sum_{\mathbf{d}_\mathrm{B}\in \mathcal{D}_\mathrm{B}}
    r_{\{\mathbf{d}_{\mathrm{K}} \vert\mathbf{d}_{\mathrm{B}}\}}(t_2,t_3)
\end{equation}
where 
\begin{align}
\label{eq:local_Rketbra_alex}
    r_{\{\mathbf{d}_\mathrm{K} \vert\mathbf{d}_\mathrm{B}\}}(t_2,t_3)
    =\left\langle\hat{V}^{-}_{R}(t_2+t_3)\hat{\sigma}^{+}_{\mathbf{d}_\mathrm{K}}\hat{\rho}(0)\hat{\sigma}_{\mathbf{d}_\mathrm{B}}^{-}\hat{V}^{+}_{R}(t_2)\right\rangle 
\end{align}
is the local response function and $\mathcal{D}_K$ and $\mathcal{D}_B$ denote decompositions of the bra and ket contributions, respectively.
Note that $\hat{\sigma}^{-}_{\mathbf{d}_\mathrm{B}}$ and $\hat{\sigma}^{+}_{\mathbf{d}_\mathrm{K}}$ act on the bra and ket sides of the initial density matrix, respectively. Because of the linearity of the Fourier transform the spectrum Eq.~(\ref{Eq:FourierTransformR}) can be written as 
\begin{align}
    \mathdutchcal{F}(\omega_R,t_2)
    &= \sum_{\mathbf{d}_\mathrm{K}, \mathbf{d}_\mathrm{B}} \mathdutchcal{f}_{\{\mathbf{d}_\mathrm{K}\vert \mathbf{d}_\mathrm{B}\}}(\omega_R,t_2)
\end{align}

with individual contributions
\begin{align}
    &\mathdutchcal{f}_{\{\mathbf{d}_\mathrm{K}\vert \mathbf{d}_\mathrm{B}\}}(\omega_R,t_2) \\
    &\quad=4D_I^2D_R^2|E_I|^2\,\textrm{Re}\left[\int_0^\infty \!\!\!r_{\{\mathbf{d}_\mathrm{K}\vert \mathbf{d}_\mathrm{B}\}}(t_2,t_3) \mathrm{e}^{-i \omega t_3} dt_3\right]. \nonumber
\end{align}

We now discuss this decomposition in the context of the dyadic HOPS:
Each local response function Eq.~(\ref{eq:local_Rketbra_alex}) can be obtained using the Dyadic HOPS as discussed in section \ref{Sec:DadHOPS}.
However, now the $\tilde{V}_1$ and $\tilde{V}_2$ operators (cf.~Eqs.~(\ref{eq:dyadic_wf}) and~(\ref{eq:V1,V2,V3})) contain $\hat{\sigma}_{\mathbf{d}_\mathrm{B}}^+$ and $
        \hat{\sigma}_{\mathbf{d}_\mathrm{K}}^+$, respectively (instead of $\hat{V}^+_I$).
Then the local response function can be obtained as an ensemble average
\begin{align}
\label{Eq:Response_Func_braket_ISD_alex}
\begin{split}
    r_{\{\mathbf{d}_\mathrm{K}\vert \mathbf{d}_\mathrm{B}\}}(t_2,t_3)= \mathcal{M}_{\mathbf{z}}\big\{ \mathbb{r}_{\{\mathbf{d}_\mathrm{K}\vert \mathbf{d}_\mathrm{B}\}}(t_2,t_3;\mathbf{z^*}) \big\} 
    \end{split}
\end{align}
over
\begin{align}
    \mathbb{r}&_{\{\mathbf{d}_\mathrm{K}\vert \mathbf{d}_\mathrm{B}\}}(t_2,t_3;\mathbf{z^*})\\
    &=I(t_2;\mathbf{z^*})\bra{\tilde{\psi}_{\{\mathbf{d}_\mathrm{K}\vert\mathbf{d}_\mathrm{B}\}}(t_2,t_3;\mathbf{z^*})}\tilde{F}\ket{\tilde{\psi}_{\{\mathbf{d}_\mathrm{K}\vert \mathbf{d}_\mathrm{B}\}}(t_2,t_3;\mathbf{z^*})}\nonumber
\end{align}
the local response function calculated from a single wave function trajectory 
\begin{equation}
        \ket{\tilde{\psi}_{\{\mathbf{d}_\mathrm{B}|\mathbf{d}_\mathrm{K}\}}(t_2,t_3;\mathbf{z^*})}= \tilde{G}(t_3;\mathbf{z^*})\tilde{V}_3\tilde{G}(t_2;\mathbf{z^*}) \ket{\tilde{\psi}_{\{\mathbf{d}_\mathrm{B}|\mathbf{d}_\mathrm{K}\}}(0)}
\end{equation}
with an initial state 
\begin{equation}
    \ket{\tilde{\psi}_{\{\mathbf{d}_\mathrm{B}|\mathbf{d}_\mathrm{K}\}}(0)} = \frac{1}{\sqrt{2}} \begin{pmatrix}\hat{\sigma}_{\mathbf{d}_\mathrm{B}}^+\ket{g}\\
  \hat{\sigma}_{\mathbf{d}_\mathrm{K}}^+\ket{g}
        \end{pmatrix}.
\end{equation}
Normalization is performed as described above in section \ref{Sec:DadHOPS}.

The significance of  Eq.~(\ref{Eq:Response_Func_braket_ISD_alex}) is that it allows us to calculated the response function as a sum of trajectories that are initially localized on small clusters of molecules, in the spirit of the localization property of the HOPS equation.

\begin{figure}
\includegraphics[width=1\linewidth]{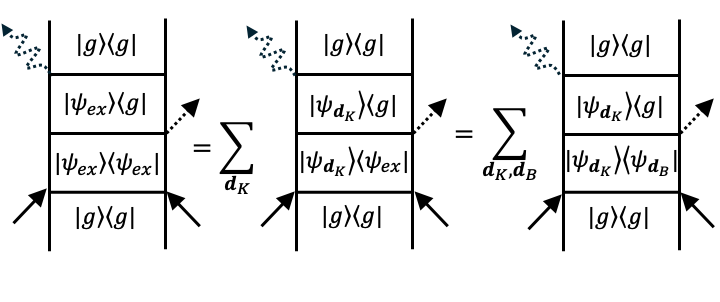}
\caption{Double-sided Feynman diagram representation of EOD.}\label{Fig:FD_EOD}
\end{figure}

\begin{figure}
\includegraphics[width=1\linewidth]{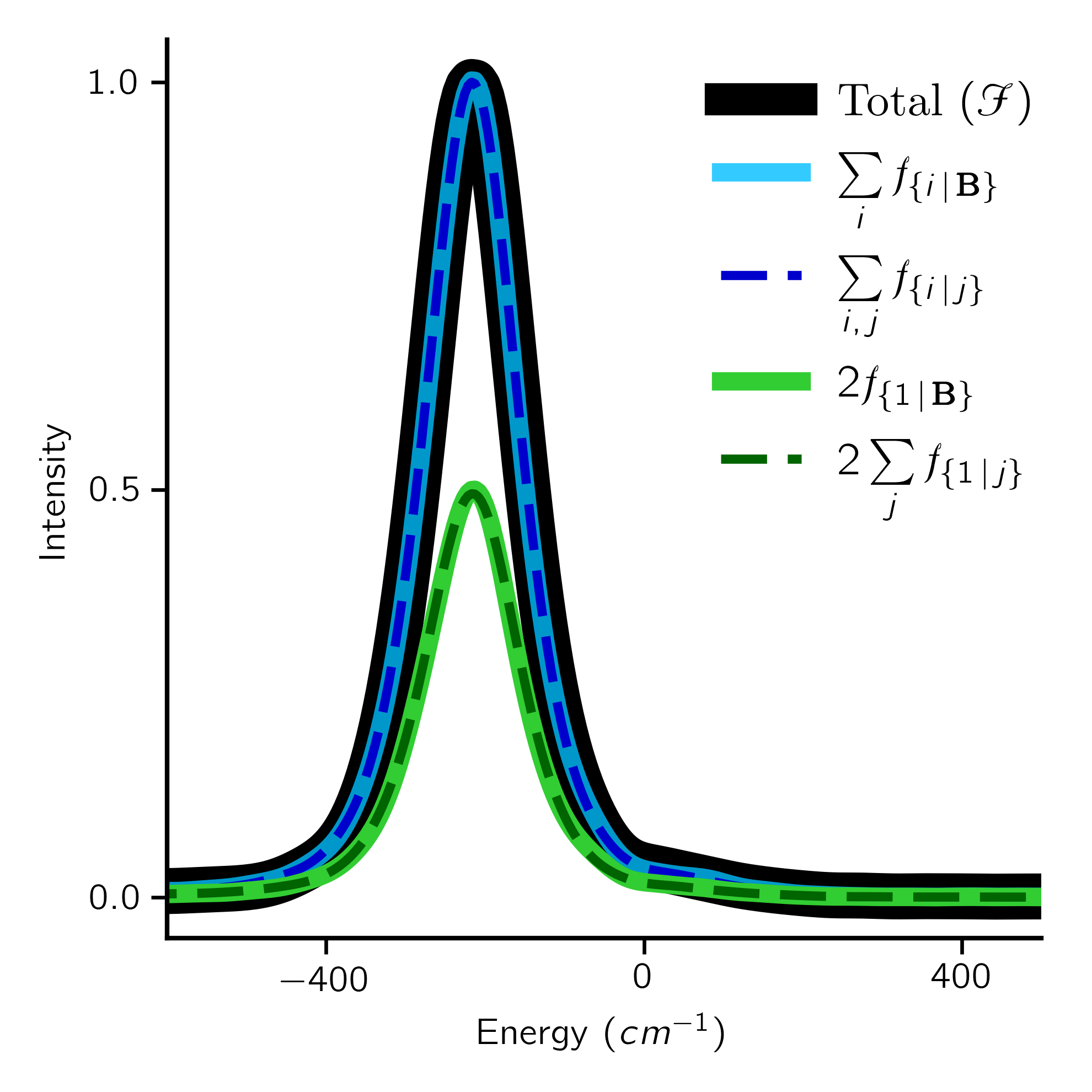}
\caption{The total fluorescence spectrum can be reconstructed from local contributions. The fluorescence spectrum for a 4-site chain (black) is reconstructed by summing over local contributions, where a single site is initially excited for the ket state (light blue) and for both the bra and ket states (dark blue dashed). The edge contribution (lime green) is reconstructed from the contributions of the bra state decomposition (dark green dashed). Parameters: $J_{n,n\pm1}=-100 \textrm{ cm}^{-1}$, $\lambda = 35 \textrm{ cm}^{-1}$, $\gamma=50 \textrm{ cm}^{-1}$, $T=295 \textrm{ K}$, $t_2 = 1000 \textrm{ fs}$, $t_3^{max} = 500 \textrm{ fs}$, and $\omega_{res} = 1.25 \times10^{-4} \,\textrm{fs}^{-1}$. Convergence parameters are given in Table \ref{tab:CS}.}\label{Fig:EOD}
\end{figure}

\subsubsection{Example: homogeneous, linear chain }
To demonstrate that the EOD in combination with the dyadic HOPS formalism reproduces the total spectrum, we perform calculations on a 4-site ($N = 4$) homogeneous linear chain ($E_n = 0$) with nearest-neighbor couplings ($J_{n,n\pm1} = -100 \textrm{ cm}^{-1}$).  We first perform EOD only for the ket states such that $\mathcal{D}_K=\{\{1\},\{2\},\{3\},\{4\}\}$ and $\mathcal{D}_B=\{1,2,3,4\}$. In this case, the total response function can be reconstructed by a sum of four single-site initial conditions 
\begin{align}
   R(t_2,t_3) &= r_{\{1 \vert \mathrm{\mathbf{B}}\}}(t_2,t_3)+r_{\{2 \vert \mathrm{\mathbf{B}}\}}(t_2,t_3)\nonumber\\
   &+r_{\{3 \vert \mathrm{\mathbf{B}}\}}(t_2,t_3)+r_{\{4 \vert \mathrm{\mathbf{B}}\}}(t_2,t_3)
\end{align}

The symmetry of the system gives rise to two pairs of equivalent contributions to this decomposition: the edge terms ($\{1\}$ and $\{4\}$) and the inner terms ($\{2\}$ and $\{3\}$). 
Therefore, we can also write the ket EOD contributions for the response function as
\begin{equation}
    R(t_2,t_3) = r_{\mathrm{edge}}(t_2,t_3) + r_{\mathrm{inner}}(t_2,t_3).\label{eq:response_decomp_ket_inner_edge}
\end{equation}
with 
\begin{align}
    r_{\mathrm{edge}}(t_2,t_3) &= r_{\{1 \vert \mathrm{\mathbf{B}}\}}(t_2,t_3) +r_{\{4 \vert \mathrm{\mathbf{B}}\}}(t_2,t_3) \label{eq:r_edge} \\    
    &= 2 r_{\{1 \vert \mathrm{\mathbf{B}}\}}(t_2,t_3)\nonumber
\end{align}    
and 
\begin{align}
    r_{\mathrm{inner}}(t_2,t_3) &= r_{\{2 \vert \mathrm{\mathbf{B}}\}}(t_2,t_3) + r_{\{3 \vert \mathrm{\mathbf{B}}\}}(t_2,t_3)\label{eq:r_inner} \\ 
    &= 2 r_{\{2 \vert \mathrm{\mathbf{B}}\}}(t_2,t_3)\nonumber
\end{align}  
where in the second lines we have taken the symmetry of the system into account.
The set $\mathbf{B}$ appearing in the above equations is used to indicate the sole element in the N-site decomposition condition $\mathcal{D}_{\textrm{all}}$.

Fig.~\ref{Fig:EOD} demonstrates the reproduction of a normalized (with respect to the peak height of $\mathdutchcal{F}$$(\omega_R, t_2)$) fluorescence spectra using the EOD: the total spectrum (black line) obtained from Eq.~\eqref{Eq:Generic_Dyadic_R} and Eq.~\eqref{eq:dyadic_wf} agrees with the reconstructed spectrum (Eq.~\eqref{eq:response_decomp_ket_inner_edge}) from the combination of the inner and edge contributions (cyan line). By noting that the total response function can be reconstructed from the decomposition of both excitation operators such that
\begin{equation}
    R(t_2,t_3) = \sum_{j=1}^4\sum_{i=1}^4 r_{\{i \vert j\}}(t_2,t_3), 
\end{equation}
we can again make use of symmetry to reconstruct the total response function
\begin{equation}
    R(t_2,t_3) = \sum_{j=1}^4 2r_{\{1 \vert j\}}(t_2,t_3) +\sum_{j=1}^4 2r_{\{2 \vert j\}}(t_2,t_3), \label{eq:response_decomp_ket_bra}
\end{equation}
where the first and second terms recreate their counterparts from Eqs. \eqref{eq:r_edge} and \eqref{eq:r_inner}, respectively. 
Accordingly, we find that the total response function (Fig.~\ref{Fig:EOD}, black line) can be reproduced by summing over the doubled local components $r_{\{1\vert j\}}$ and $r_{\{2\vert j\}}$ (Fig.~\ref{Fig:EOD}, dark blue line). We also show that the edge component (Fig.~\ref{Fig:EOD}, lime green line) can be reconstructed by summing over the doubled $r_{\{1 \vert j\}}$ components (Fig.~\ref{Fig:EOD}, dark green line).

\begin{figure}
\includegraphics[width=1\linewidth]{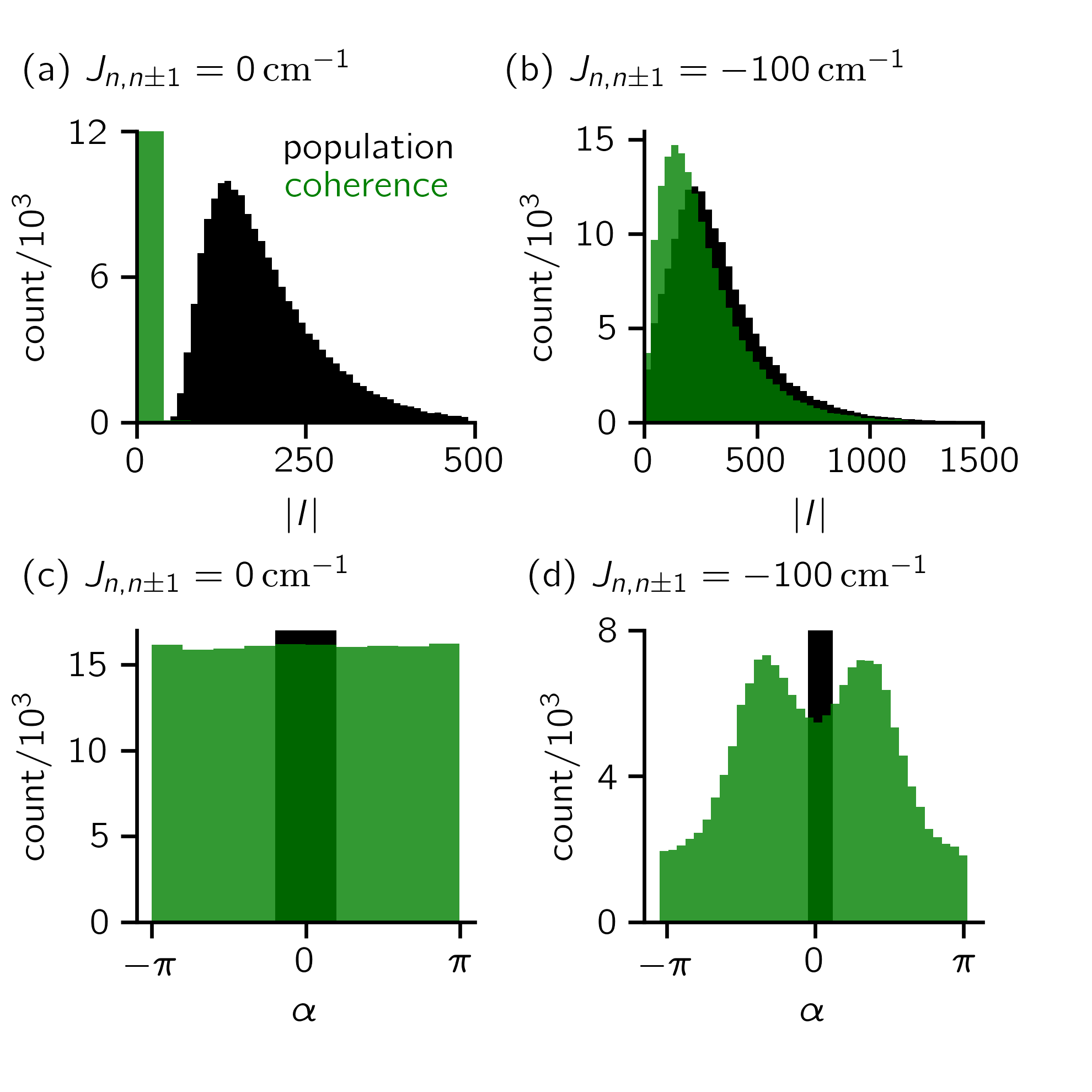}
\caption{Population contributions are the most relevant contributions, independent of electronic coupling. 
 The spread (over 100000 trajectories) in the integrated amplitude (over $t_3$) is plotted in black for $\mathbb{r}_{\{1\vert1\}}(t_2,t_3)$ (population contribution) and in green for $\mathbb{r}_{\{1\vert2\}}(t_2,t_3)$ (coherence contributions) in panels (a) and (b) for  $ J_{n,n\pm1}=0\,\textrm{cm}^{-1}$ and $ J_{n,n\pm1}=-100\,\textrm{cm}^{-1}$, respectively. We used a larger bin size for $\mathbb{r}_{\{1\vert2\}}(t_2,t_3)$ in panel (a) to enhance the visibility of the histogram. The phase spread at $t_2$ is plotted in black for $\mathbb{r}_{\{1\vert1\}}(t_2,t_3)$ and in green for $\mathbb{r}_{\{1\vert2\}}(t_2,t_3)$ in panels (c) and (d) for  $ J_{n,n\pm1}=0$ and $ J_{n,n\pm1}=-100~\textrm{cm}^{-1}$, respectively. We used a larger bin size for $\mathbb{r}_{\{1\vert1\}}(t_2,t_3)$ in panel (d) to enhance the visibility of the histogram. Parameters: $\lambda = 35~\textrm{cm}^{-1}$, $\gamma=50~\textrm{cm}^{-1}$, $T=295~\textrm{K}$, $t_2=200\,\textrm{fs}$, and $t_3^{\textrm{max}}=500\,\textrm{fs}$. 
 The convergence parameters are given in Table~\ref{tab:CS}.}\label{Fig:fig4}
\end{figure}

\subsection{Monte Carlo Sampling with Dyadic Adaptive HOPS (DadHOPS)}

Here, we demonstrate a size-invariant (i.e.~independent of the number of molecules $N$, $N^0$) scaling algorithm for fluorescence calculations that combines Dyadic adaptive HOPS (DadHOPS)\cite{Gera2023} with the local response function formulation presented in Section \ref{Sec:EOD}. Size-invariant scaling fluorescence calculations require an algorithm that is: first, capable  of calculating individual trajectories with size-invariant scaling, and, second, requires a number of trajectories that remains independent of system size to achieve a given accuracy.

DadHOPS achieves size-invariant scaling for individual trajectories using a time-evolving basis set. The adaptive HOPS (adHOPS) algorithm \cite{Varvelo2021} dynamically constructs a new time-dependent reduced basis ($\mathbb{A}_t \bigotimes \mathbb{S}_t$) of auxiliary wave functions ($\mathbb{A}_t \in \mathbb{A}$) and pigment states ($\mathbb{S}_t \in \mathbb{S}$) every $u_s$ time steps, while ensuring a user-defined bound on the derivative error $\delta = \sqrt{\delta_\mathrm{A}^2 + \delta_\mathrm{S}^2}$. 
The adHOPS algorithm was originally developed to simulate exciton dynamics in molecular materials, such as photosynthetic aggregates,\cite{Varvelo2023} and was first extended to the DadHOPS algorithm to simulate absorption spectra.\cite{Gera2023} Recent work has improved the numerical performance of the adHOPS algorithm by introducing an additional convergence parameter $f_\mathrm{dis}$ which efficiently filters small derivative error to limit the computational expense of constructing the reduced auxiliary basis ($\mathbb{A}_t$).\cite{Citty2024}

Efficient Monte Carlo sampling of the total response function requires simultaneously sampling the bra/ket EOD and the noise trajectories that define the HOPS propagation ($\mathbf{z^*}$). The local response function can be calculated as the average over a finite set of $M$ trajectories ($\mathcal{M}_\mathbf{z}\rightarrow \frac{1}{M}\sum_{\mathbf{z}}^{M}$) such that
\begin{align}
\begin{split}
    r_{\{\mathbf{d}_\mathrm{K}\vert \mathbf{d}_\mathrm{B}\}}(t_2,t_3)&= \mathcal{M}_{\mathbf{z}}\big\{ \mathbb{r}_{\{\mathbf{d}_\mathrm{K}\vert \mathbf{d}_\mathrm{B}\}}(t_2,t_3;\mathbf{z^*}) \big\}\\
    &=\frac{1}{M_{\{\mathbf{d}_\mathrm{K}\vert \mathbf{d}_\mathrm{B}\}}}\sum_{\mathbf{z}}^{M_{\{\mathbf{d}_\mathrm{K}\vert \mathbf{d}_\mathrm{B}\}}} \mathbb{r}_{\{\mathbf{d}_\mathrm{K}\vert \mathbf{d}_\mathrm{B}\}}(t_2,t_3;\mathbf{z^*}).
    \end{split}
\end{align}
In this notation, the total response function is given by 
\begin{eqnarray}
    R(t_2,t_3)&=\sum_{\mathbf{d}_\mathrm{K}\in \mathcal{D}_\mathrm{K}} 
    \sum_{\mathbf{d}_\mathrm{B}\in \mathcal{D}_\mathrm{B}}\left[
     \frac{1}{M_{\{\mathbf{d}_\mathrm{K}\vert \mathbf{d}_\mathrm{B}\}}}\right.\nonumber\\
     &\left.\times\sum_{\mathbf{z}}^{M_{\{\mathbf{d}_\mathrm{K}\vert \mathbf{d}_\mathrm{B}\}}} \mathbb{r}_{\{\mathbf{d}_\mathrm{K}\vert \mathbf{d}_\mathrm{B}\}}(t_2,t_3;\mathbf{z^*})\right]\label{Eq:Response_Unbaised_MC}
\end{eqnarray}
which, naively, appears to require converging $N_\mathrm{D}=N_\mathrm{bra}\cdot N_\mathrm{ket}$ local response functions, where $N_\mathrm{bra}$ and $N_\mathrm{ket}$ represent the number of clusters in the bra and ket decompositions, respectively. However, the ensemble average over the noise trajectories ($\mathbf{z^*}$) is independent of the initial conditions ($\mathbf{d}_\mathrm{K} ,\, \mathbf{d}_\mathrm{B}$), which allows for an efficient Monte Carlo scheme where the noise and initial conditions are sampled simultaneously to converge the total response function. If the bra and ket cluster decompositions consist of same-size clusters, then unbiased sampling leads to $N_{\mathbf{d}_\mathrm{K},\mathbf{d}_\mathrm{B},\mathbf{z^*}} = N_{\mathrm{traj}}/N_D$ noise trajectories per initial condition, and the total response function can be calculated as
\begin{align}
\label{eq:R(t)_total_montecarlo}
R(t_2,t_3)=  \frac{N_D}{N_{\textrm{traj}}} \sum_{(\mathbf{d}_\mathrm{K},\mathbf{d}_\mathrm{B}, \mathbf{z^*})} \mathbb{r}_{\{\mathbf{d}_\mathrm{K}\vert \mathbf{d}_\mathrm{B}\}}(t_2,t_3;\mathbf{z^*})
\end{align}
where we Monte Carlo sample $N_{\mathrm{traj}}$ sets of $(\mathbf{d}_\mathrm{K},\mathbf{d}_\mathrm{B}, \mathbf{z^*})$ to calculate the total correlation function.

\subsubsection{Unbiased Monte Carlo sampling is inefficient for large aggregates}
Unbiased Monte Carlo sampling of the bra and ket excitation operators is inefficient for large systems because most local response functions have a near zero amplitude as we discuss now. 

In dyadic HOPS, a local response function ($r_{\{\mathbf{d}_\mathrm{K}\vert\mathbf{d}_\mathrm{B}\}}(t_2,t_3)$) can decay in amplitude either due to decoherence (i.e., pure dephasing) where different components of a wave function lose their phase relationship, leading to a loss of amplitude for each individual realization of the response function or ensemble dephasing where different realizations of the response function undergo destructive interference. In the limit of two uncoupled (i.e., infinitely separated) molecules, the population contributions ($\mathbf{d}_\mathrm{K} = \mathbf{d}_\mathrm{B}$) are not limited by decoherence (black, Fig.~\ref{Fig:fig4}(a)) and show negligible spread in phase (black, Fig.~\ref{Fig:fig4}(c)). On the other hand, the coherence contributions ($\mathbf{d}_\mathrm{K} \neq \mathbf{d}_\mathrm{B}$) have small amplitudes associated with decoherence (green, Fig.~\ref{Fig:fig4}(a)) and a wide phase distribution (green, Fig.~\ref{Fig:fig4}(c)) leading to ensemble dephasing which further suppresses their average magnitude. When molecules have an electronic coupling larger than the reorganization energy of the thermal environment, there is limited decoherence (Fig.~\ref{Fig:fig4}(b)) in both the population and the coherence contributions. However, a wide spread in phase factors is observed for coherence contributions (green, Fig.~\ref{Fig:fig4}(d)), leading to increased ensemble dephasing for these response functions. As expected, for coherence contributions the extent of ensemble dephasing increases with increasing $t_2$ (data not shown).  

As a result, in extended aggregates there is a large ($N^2$) number of local response functions with negligible contributions, and a naive sampling of initial excitation conditions that averages over all of these (mostly zero) contributions will require a total number of trajectories that grows with system size.

\subsubsection{Efficient Monte Carlo sampling in a linear chain}\label{Sec:Eff_MC_sampling}

Achieving size-invariant scaling fluorescence calculations with DadHOPS requires an efficient sampling scheme for the $N_\mathrm{bra}\cdot N_\mathrm{ket}$ contributions to the bra and ket EOD. Recognizing that the presence of ensemble dephasing means that trajectories with non-zero amplitude can contribute to local response functions with net zero amplitude, we employ an algorithm that allows us to use ensemble-level insights to determine our sampling approach. First, we randomly select a ket excitation operator, \( \hat{\sigma}^{+}_{\mathbf{d}_\mathrm{K}} \), to initially excite the cluster \( \mathbf{d}_\mathrm{K} \). In this work, we use a single-site decomposition for the ket contributions; however, multiple sites can also be included within a single cluster \( \mathbf{d}_\mathrm{K} \). Care should be taken, however, since clusters much larger than the exciton delocalization extent will lead to unnecessarily large basis sets in the early time component of DadHOPS trajectories.
 Second, we select the bra side contributions to include $\mathbf{d}'_\mathrm{B}$ vs. exclude $\mathbf{d}''_\mathrm{B}$ (i.e, bra side EOD is defined by $\mathcal{D}_{\mathbf{B}}=\{\mathbf{d}'_\mathrm{B}, \mathbf{d}''_\mathrm{B}\}$) . 
Formally, this is equivalent to decomposing the $r_{\{\mathbf{d}_\mathrm{K}\vert \mathrm{\mathbf{B}}\}}(t_2,t_3)$ term into a sum over two contributions
\begin{align}
r_{\{\mathbf{d}_\mathrm{K}\vert \mathrm{\mathbf{B}}\}}(t_2,t_3) &=\mathcal{M}_{\mathbf{z^*}}\{ \mathbb{r}_{\{\mathbf{d}_\mathrm{K}\vert\mathbf{d}_\mathrm{B}'\}}(t_2,t_3;\mathbf{z^*}) \}\nonumber\\
&+\mathcal{M}_{\mathbf{z^*}}\{\mathbb{r}_{\{\mathbf{d}_\mathrm{K}\vert\mathbf{d}_\mathrm{B}''\}}(t_2,t_3;\mathbf{z^*})\rangle\}\label{Eq:full_EOD_divided}
\end{align}
where we calculate the contribution from $\mathbf{d}_\mathrm{B}'$ explicitly and approximate the contribution from  $\mathbf{d}_\mathrm{B}''$ as zero. For a nearest-neighbor coupled linear chain, the most important bra excitations to include in $\mathbf{d}_\mathrm{B}'$ will be the sites closest to the initial ket excitations. For a more complicated Hamiltonian, Appendix~\ref{AppA} proposes a general algorithm for selecting the most important bra excitations to include in the set $\mathbf{d}_\mathrm{B}'$.

\begin{figure}
\includegraphics[width=1\linewidth]{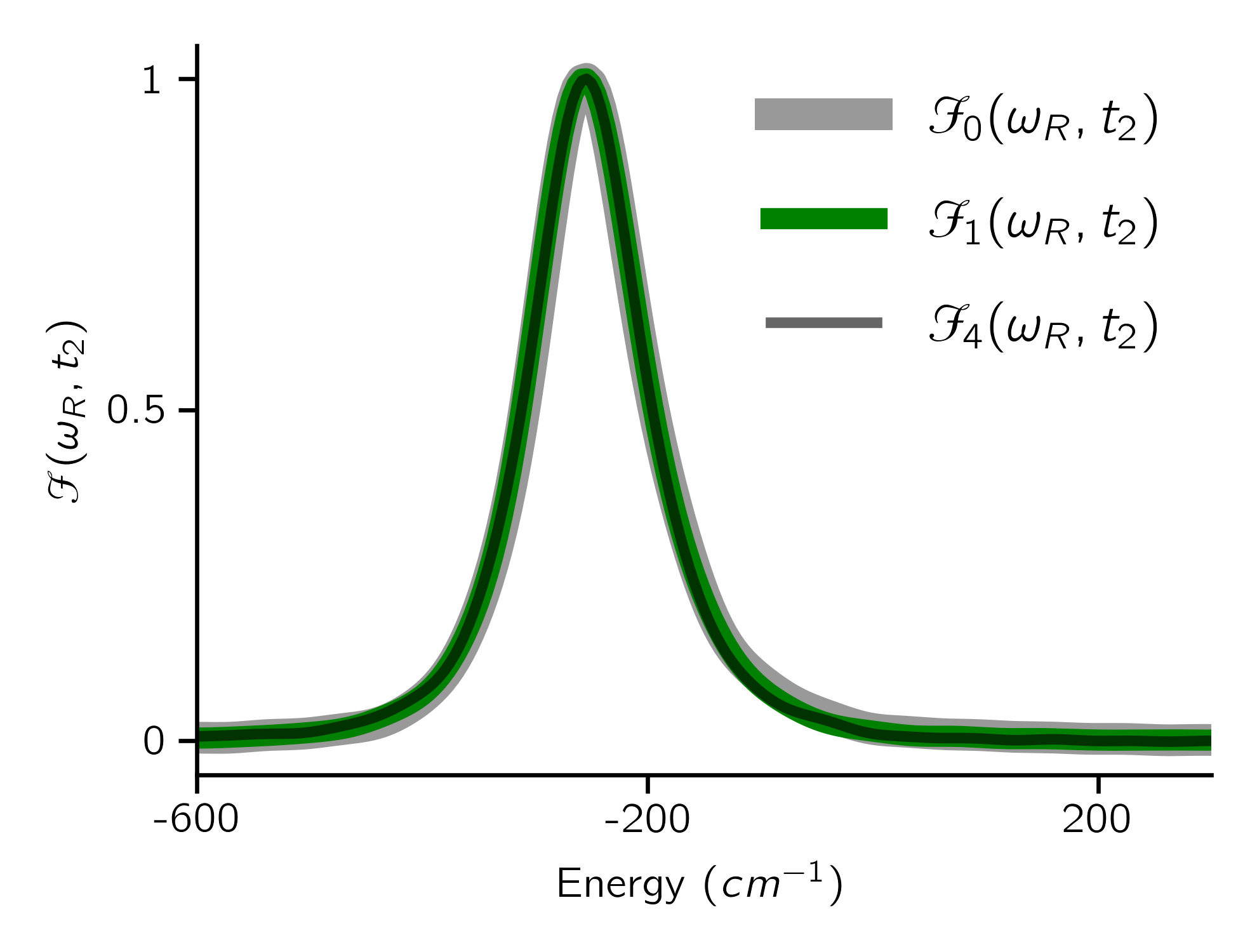}
\caption{
Population contributions are sufficient to determine the normalized fluorescence spectrum. 
 Normalized spectrum for a 100-site chain for different values of \( n \) in \( \mathdutchcal{F}_n \), with \( \mathdutchcal{F}_0(\omega_R,t_2) \) (gray), \( \mathdutchcal{F}_1(\omega_R,t_2) \) (green), and \( \mathdutchcal{F}_4(\omega_R,t_2) \) (black).
Parameters: $ J_{n,n\pm1}=-100 \textrm{ cm}^{-1}$, $\lambda = 35 \textrm{ cm}^{-1}$, $\gamma=50 \textrm{ cm}^{-1}$, $T=295 \textrm{ K}$, $t_2=400\,\textrm{fs}$, $t_3^{\textrm{max}}=500\,\textrm{fs}$, and $\omega_{\textrm{res}}=1.25\times10^{-4} \,\textrm{fs}^{-1}$. The convergence parameters are given in Table~\ref{tab:CS}.} \label{Fig:fig6}
\end{figure}

For linear chains with nearest-neighbor coupling, the population contributions alone are sufficient to reproduce the normalized fluorescence spectra.
In Fig.~\ref{Fig:fig6}, we Monte Carlo sample $N_{\textrm{traj}}$ pairs of noise $(\mathbf{z^*})$ and single-site clusters for the ket site $(i)$ and plot the normalized (with respect to peak height) fluorescence spectrum for different choices of $\mathbf{d}_\mathrm{B}'$ using
\begin{align}
\mathdutchcal{F}_{n}(\omega_R,t_2)
&=\frac{4D_I^2D_R^2|E_I|^2}{N_{\textrm{traj}}}\\
&\times \int_0^\infty\,dt_3\,\mathrm{e}^{-i \omega_R t_3}\sum_{(i, \mathbf{z^*})}\textrm{Re}\left[ \mathbb{r}_{\{i\vert\mathbf{d}_i^n\}}(t_2,t_3;\mathbf{z^*}) \right]\nonumber 
\end{align}
where $\mathbf{d}_i^n = \{ k \in \mathbb{Z} \mid k = i \pm j, \; 0 \leq k \leq N, \; 0 \leq j \leq n \}$ is the set of bra sites within $n$ steps of the randomly selected ket initial site $i$. The agreement between $n=0$ (gray) with $n=4$ (black) in Fig.~\ref{Fig:fig6} demonstrates that sampling only the population contributions captures the normalized fluorescence lineshape.

Most experiments only measure the normalized fluorescence spectrum, but we extend our study to the absolute spectrum in Appendix \ref{App:Abs_spectrum} and demonstrate convergence in sampling the initial bra state, allowing $\mathbf{d}'_{B}$ to be truncated to a reasonable cluster size $l_{b} \ll N$, to avoid unnecessary sampling. We expect that the importance of coherence terms will depend on both the Hamiltonian parameters and the population time of the simulation. Nevertheless, simulating normalized spectra has substantial numerical advantages for the method presented here.

\subsubsection{Size-invariant scaling fluorescence calculations}

\begin{figure}
\includegraphics[width=1\linewidth]{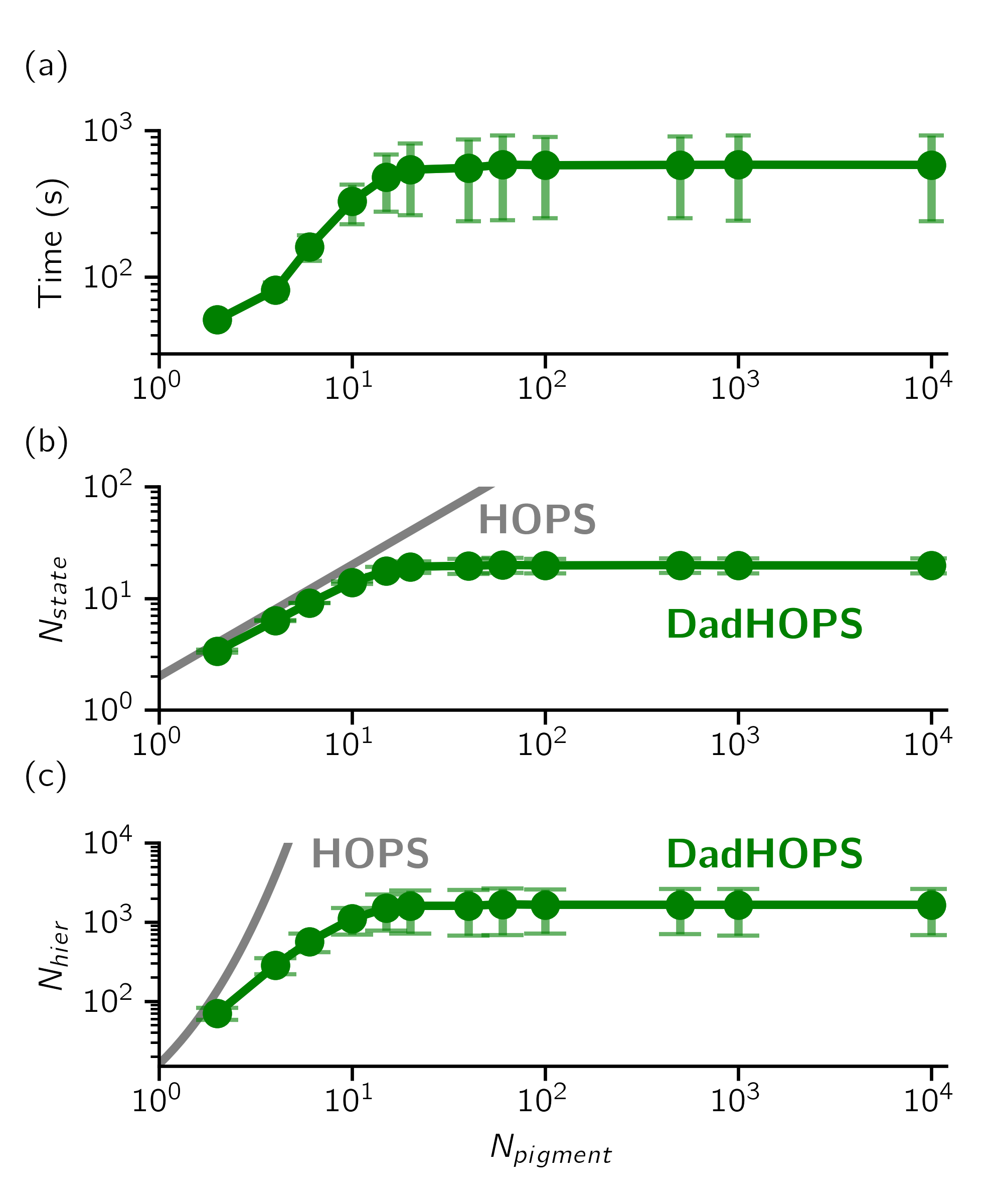}
\caption{DadHOPS fluorescence calculations have size-invariant scaling for sufficiently large aggregates. The green circles mark (a) average CPU time, (b) number of elements in the adaptive state basis and (c) average number of elements in the adaptive auxiliary basis required for each chain length. Here, we initially excite only the first site for both the ket and bra states. For each data point the average and standard deviation were determined using 2000 trajectories. Parameters: $ J_{n,n\pm1}=-100 \textrm{ cm}^{-1}$, $\lambda = 35 \textrm{ cm}^{-1}$, $\gamma=50 \textrm{ cm}^{-1}$,  $T=295 \textrm{K}$, $t_2=400\,\textrm{fs}$, and $t_3^{\textrm{max}}=500\,\textrm{fs}$. The convergence parameters are given in Table~\ref{tab:CS}.}\label{Fig:Size_invariance}
\end{figure}

\begin{figure}
\includegraphics[width=1\linewidth]{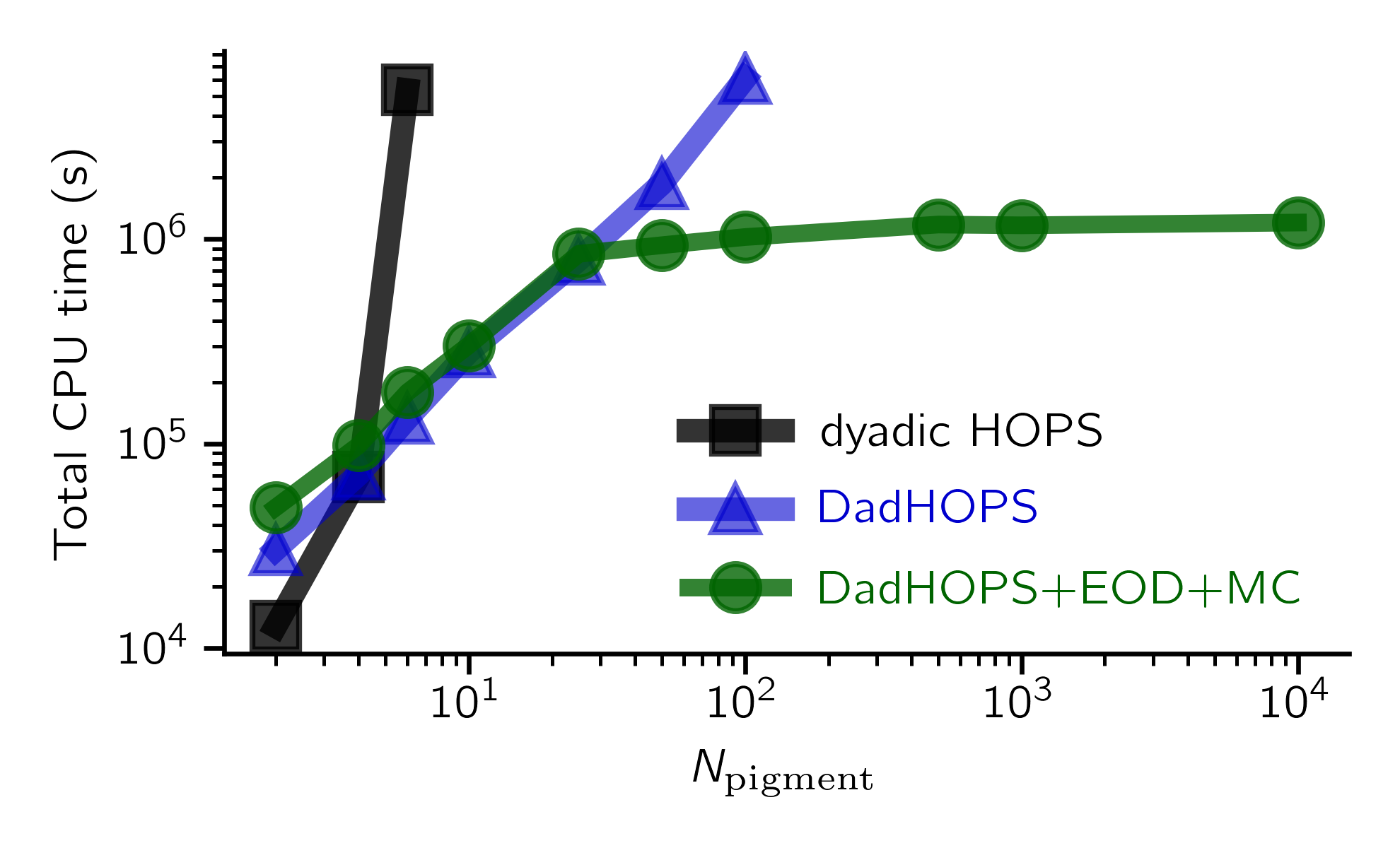}
\caption{Combining DadHOPS with EOD and MC sampling yields size-invariant scaling fluorescence calculations for sufficiently large molecular aggregates. Green circles represent the total CPU time for DadHOPS with both EOD and MC sampling, blue triangles denote DadHOPS without decomposition, and black squares indicate dyadic HOPS. Each data point reflects the total time required to compute 1000 trajectories.
 Parameters: $J_{n,n\pm1}=-100 \textrm{ cm}^{-1}$, $\lambda = 35 \textrm{ cm}^{-1}$, $\gamma=50 \textrm{ cm}^{-1}$, $T=295 \textrm{K}$, $t_2=400\,\textrm{fs}$, and $t_3^{\textrm{max}}=500\,\textrm{fs}$. The convergence parameters are given in Table~\ref{tab:CS}.}\label{Fig:method_scan}
\end{figure}

DadHOPS in combination with EOD becomes size-invariant ($N^0$) in sufficiently large systems. 
To demonstrate size-invariance, we use the model introduced in Sec.~\ref{Sec:EOD} for linear chains with different number of pigments $N$. Fig.~\ref{Fig:Size_invariance}(a) shows the average (over 2000 trajectories) CPU time needed to simulate fluorescence as a function of the number of pigments. The average time starts to stabilize at $N=20$ and becomes size-invariant from $N=50$ onwards. Fig.~\ref{Fig:Size_invariance}(b) and Fig.~\ref{Fig:Size_invariance}(c) plot the change in system basis size with increasing chain length. The DadHOPS adaptive state and auxiliary basis sizes (green circles) stabilize and cease to grow with increasing chain length, whereas in HOPS (grey line) the full basis of the dyadic construct scales catastrophically. \\
To further assess the computational efficiency of each method, we present the total CPU time required to run 1000 trajectories using different approaches in Fig.~\ref{Fig:method_scan}. Dyadic HOPS (black squares) is the most efficient for small systems such as dimers and tetramers, but its computational cost grows rapidly with system size, becoming impractical for larger chains. With DadHOPS (blue squares), the computational time increases more steadily; however, for longer chains, we encounter growing demands in both CPU time and memory. When DadHOPS is combined with Monte Carlo sampled pathways defined by the Excitation Operator Decomposition using a single-site decomposition strategy (green circles), the total CPU time is consistent with DadHOPS (blue triangles) up to a chain length of 25, after which it plateaus— demonstrating the onset of size-invariant performance.


\section{Application: $N$ dependent Stokes Shift and Delocalization extent in J-Aggregates}
\label{sec:Application}
Here, we demonstrate that the ensemble average of the inverse participation ratio (IPR) of HOPS wave functions, a measure of delocalization across molecules, reproduces the coherence number extracted from fluorescence spectra.
In this section we describe the system-bath coupling using a Drude-Lorentz spectral density (Eq~\eqref{Eq:DL_spectral}) in combination with an underdamped Brownian oscillator: 
\begin{equation}
W^{BO}_n(\omega) = \frac{2\lambda \gamma \omega\chi^2}{(\omega^2-\chi^2)+\omega^2 \gamma^2}
\end{equation}
The corresponding bath correlation function will be a sum of two exponentials, and we neglect the non-resonant term and keep the resonant one to model our thermal bath, given by
\begin{equation}
    \alpha_n(t) = \frac{\lambda\chi^2}{2\xi}\left(\coth\left(\frac{\beta(\xi-i\gamma/2)}{2}+1\right)\right) e^{-\left(\frac{\gamma}{2}+i\xi\right) t/\hbar}\label{Eq:BO_bath}
\end{equation}
 where $\xi=\sqrt{\chi^2-\gamma^2/4}$ and the parameters used in this calculations are $g=(2.05\times10^4-2.5\times10^3i)\, \textrm{cm}^{-2}$, $\gamma=50\, \textrm{cm}^{-1}$, $\lambda=50\, \textrm{cm}^{-1}$ corresponding to an over-damped Drude Lorentz mode (Eq.~\eqref{Eq:DL_bath}), and  under-damped mode (Eq.~\eqref{Eq:BO_bath}) with $\lambda=1023\,\textrm{cm}^{-1}$, $\chi=1550\,\textrm{cm}^{-1}$, $\gamma=200\, \textrm{cm}^{-1}$,  and $T=298$ K.\\
Fluorescence spectroscopy, and particularly super-radiance, can report on the exciton delocalization extent (also called the coherence number) in molecular aggregates.\cite{Spano2007, Jimenez1996, Zhao1999, Sung2015} For a periodic J-aggregate, assuming the dynamic disorder from vibrational fluctuations can be described by a static disorder on site energies and that electronic delocalization occurs in the vibrational ground-state, Spano and coworkers\cite{Spano2010} have reported a relationship between coherence number $(N_c)$ and fluorescence spectra,  
\begin{equation}
    N_{\textrm{c}}=\frac{I_{0-0}}{I_{0-1}}S^2\label{Eq:Spano_N_coh}
\end{equation}
where $I_{0-0}$ is the intensity of the $0-0$ vibronic peak, $I_{0-1}$ is intensity of the $0-1$ vibronic side band, and $S$ is the Huang-Rhys (HR) factor of the high-frequency vibration.
The delocalization extent can also be quantified through the HOPS wave functions, which are pure states, using  
\begin{equation}
    \mathrm{IPR}=\mathcal{M}_{\mathbf{z}}\left[\frac{1}{\sum_n|c_n(t_2; \mathbf{z^*})|^4}\right]\label{Eq:IPR}
\end{equation}
where $c_n(t_2; \mathbf{z^*})$ is the coefficient of the $n^{\textrm{th}}$ pigment in the physical wave function $\ket{\tilde{\psi}^{\vec{(0)}}(t_{2};\mathbf{z^*})}=\sum_n c_n(t_2;\mathbf{z^*} )\ket{n} $ calculated at the time $t=t_{2}$, i.e., directly before the first interaction with the radiation mode. 
 
\begin{figure}
\includegraphics[width=1\linewidth]{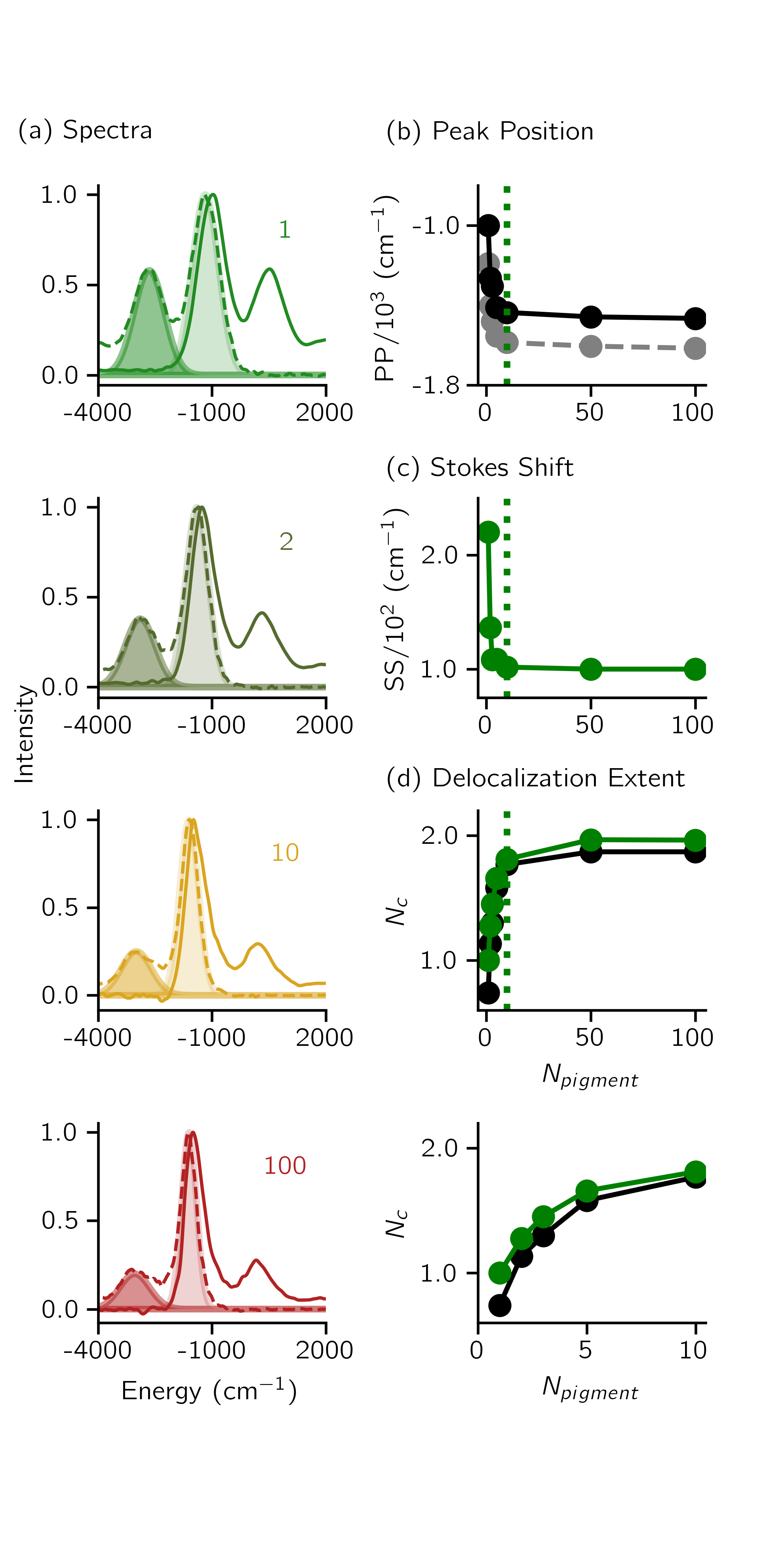}
\caption{IPR accurately reproduces the delocalization length extracted from fluorescence spectra.
  (a) Absorption (solid) and fluorescence (dashed) spectrum for monomer (green), dimer (dark green), 10-site (mustard yellow), and 100 site J-aggregates (crimson red). The shaded areas are the individually fitted Gaussian peaks for 0-0 and 0-1 features. (b) Peak position of the 0-0 peaks in absorption (black solid) and fluorescence (grey dashed). (c) Stokes shift as a function of the number of pigments. (d) Comparison of coherence number (black) and IPR (green) for number of pigments ranging from 1 to 100 (top) and 1 to 10 (bottom).  Parameters: $T=298 \textrm{K}$, $t_2=400\,\textrm{fs}$, $t^{\textrm{max}}_3=500\,\textrm{fs}$, $g_1=(2.05\times10^4-2.5\times10^3i)\, \textrm{cm}^{-2}$, $\gamma_1=50\, \textrm{cm}^{-1}$, $g_2 = 1.6\times10^6\, \textrm{cm}^{-2}$, $\gamma_2 = (100 + 1550i)\, \textrm{cm}^{-1}$,  $\omega_{\textrm{res}}=5\times10^{-4}\,\textrm{fs}^{-1}$, and $ J_{n,n\pm1}=-300\, \textrm{cm}^{-1}.$ For absorption spectra, $t_{\textrm{max}}=250\,\textrm{fs}$. The convergence parameters are given in Table~\ref{tab:CS}.}\label{Fig:fig8}
\end{figure}

Here, we compare the ensemble average HOPS IPR to the coherence number ($N_c$) extracted from fluoresence spectra for a model system that resembles the peryelene bis-imide (PBI) parameters reported in Ref.~\onlinecite{Gera2023}.
The electronic excited state energies of the individual pigments are sampled from a Gaussian distribution with a mean of 0 and a standard deviation of $300\,\textrm{cm}^{-1}$ and the electronic coupling between neighboring pigments is $-300\,\textrm{cm}^{-1}$. 
Normalized absorption (solid) and fluorescence (dashed) for a monomer, dimer, 10-site chain, and 100-site chain are shown in Fig.~\ref{Fig:fig8}(a). Fig.~\ref{Fig:fig8}(b) shows a rapid redshift in both absorption and fluorescence spectra with increasing chain length for $N_\mathrm{pigment}<10$, and a similar trend can be observed in the Stokes shift (Fig.~\ref{Fig:fig8}(c)). The central frequency shift and spectral narrowing of the 0-0 transition peak with increasing chain length highlights the entanglement between electronic and vibrational degrees of freedom—a feature accurately captured by the HOPS equation of motion. To further illustrate this effect, we present fluorescence spectra for H-aggregate analogs in Appendix \ref{App:H_agg}.

The ensemble-averaged IPR calculated with DadHOPS is found to be consistent with the coherence number $N_c$ for long chain J-aggregates. 
The coherence number in Eq.~\eqref{Eq:Spano_N_coh} is determined using the intensities of the 0-0 and 0-1 peaks, which are obtained by calculating the area under each peak after fitting them to individual Gaussian functions (shaded areas in Fig.~\ref{Fig:fig8}(a)). 
In Fig.~\ref{Fig:fig8}(d) we compare the coherence number $(N_c)$ calculated using a Huang-Rhys factor of $S=\lambda/\chi=1023/1550=0.66$ (black) with the HOPS ensemble average IPR (green). For the monomer case (Fig.~\ref{Fig:fig8}(d) bottom panel), the HOPS ensemble IPR is 1, while the coherence number is less than 1 because the analytical expression is derived\cite{Spano2010} under the assumption of periodic boundary conditions, which requires $N_\mathrm{pigment}\gg 1$. As the chain length increases, the limitation of periodic boundary conditions diminishes, and the coherence number and ensemble average IPR show remarkable agreement, with a discrepancy of less than $5\%$.

Given their close agreement, our results suggest that the HOPS ensemble average IPR is a powerful extension of the coherence number extracted from fluorescence spectra. While the coherence number is calculated assuming dynamic reorganization of the bath can be modeled as static disorder,\cite{Hestand2018} the HOPS ensemble average IPR remains physically intuitive without requiring additional approximations on the equation-of-motion. Unlike the coherence number, however, the HOPS ensemble IPR cannot be directly extracted from experimental observables. Nevertheless, we expect this to be advantageous when studying delocalization in the presence of complicated vibrational environments or as a function of temperature.

\section{Conclusions}
In this work, we introduce a size-invariant (i.e., $N^0$) algorithm for calculating fluorescence in mesoscale molecular aggregates when the measurement time exceeds the optical dephasing time. Simulating third-order pathways, such as fluorescence, in mesoscale systems has historically been challenging due to catastrophic computational scaling. Reconstructing optical response functions from Monte Carlo sampling local contributions enables size-invariant fluorescence calculations for mesoscale molecular aggregates. 
Furthermore, the agreement we have demonstrated between the coherence number extracted from fluorescence spectra and the ensemble average IPR from DadHOPS calculations underscores the effectiveness of our approach in describing the interplay between electronic delocalization and dynamic localization induced by vibrational reorganization. 
The formalism introduced here is compatible with the current adaptive HOPS (adHOPS) framework, including the recent low-temperature correction.\cite{Citty2024}  While we cannot make a general statement about computational cost, since this depends strongly on the Hamiltonian parameters, we note that the calculation time tends to scale rapidly with the number of bath modes per molecule. For instance, in the present study, including both an under- and overdamped mode increases run time by nearly an order of magnitude compared with calculations using a single overdamped mode. 

DadHOPS offers a robust computational tool for exploring absorption and fluorescence spectra in large molecular aggregates, and this work opens the door to new computational approaches for efficiently simulating spatially resolved fluorescence and non-linear (e.g., two-dimensional electronic) spectroscopy. For example, the current method is directly applicable to the recently proposed 2DFlex measurements which use an excitation-pulse pair combined with fluorescence up-conversion based read-out to extract information about excitation and emission correlation.\cite{Yang2023} In the context of two-dimensional electronic spectroscopy, which involves three primary pathways—stimulated emission (SE), ground-state bleach (GSB), and excited-state absorption (ESA)—
the current work represents an approach describing a special case of the SE pathways. Future work will extend this approach to enable efficient decomposition schemes for the GSB and ESA signals as well.

\begin{acknowledgments}
The authors thank Jacob K Lynd for editing and review.
AE acknowledges support from  the German Research Foundation (DFG) via a Heisenberg fellowship (Grant No. EI 872/10-1).
TG, AH, and DIGBR acknowledge support from the Robert A. Welch Foundation (Grant No. F-2207-20240404) and
start-up funds from the University of Texas at Austin. DIGBR and TG acknowledge the support from the U.S. National Science Foundation CAREER Award (Grant No. CHE-2341178).  AH is supported by the U.S. Department of Energy, Office of Science, Office of Advanced Scientific Computing Research, Department of Energy Computational Science Graduate Fellowship under Award Number DE-SC0025528. Computational analyses were performed using the Biomedical Research Computing Facility at UT Austin, Center for Biomedical Research Support. RRID$\#$: SCR$\_$021979. The authors also acknowledge the Texas Advanced Computing Center (TACC) at The University of Texas at Austin for providing computational resources that have contributed to the research results reported within this paper. URL: http://www.tacc.utexas.edu.
\end{acknowledgments}
\section*{Author Declarations}
\subsection*{Conflict of Interest}
The authors have no conflicts to disclose.
\subsection*{Author Contribution}
\textbf{Tarun Gera:} Methodology (equal); Software (equal); Writing-original draft (equal); writing-review \& editing (equal); Investigation (lead); Visualization (equal). \textbf{Alexia Hartzell:} Methodology (supporting); Software (equal); Writing-original-draft (supporting); writing-review \& editing (equal); Investigation (supporting); Visualization (equal).    
\textbf{Lipeng Chen:} Methodology (supporting).
\textbf{Alexander Eisfeld:} Methodology (supporting); writing-review \& editing (equal).
\textbf{Doran I. G. B. Raccah:} Methodology (equal); Software (supporting); Writing-original-draft (equal); writing-review \& editing (equal); Visualization (supporting); Project administration (lead); Supervision (lead). 
\section*{Data Availability Statement}

Simplified input, analysis, and generation scripts along with the packaged data for main text figures and the MesoHOPS source code used for these calculations are available on Zenodo.\cite{Zenodo}
\section*{Disclaimer}
This report was prepared as an account of work sponsored by an agency of 
the United States Government.  Neither the United States Government nor any agency 
thereof, nor any of their employees, makes any warranty, express or implied, or 
assumes any legal liability or responsibility for the accuracy, completeness, or usefulness 
of any information, apparatus, product, or process disclosed, or represents that its use 
would not infringe privately owned rights.  Reference herein to any specific commercial 
product, process, or service by trade name, trademark, manufacturer, or otherwise does 
not necessarily constitute or imply its endorsement, recommendation, or favoring by 
the United States Government or any agency thereof.  The views and opinions of 
authors expressed herein do not necessarily state or reflect those of the United States 
Government or any agency thereof.
\newpage
\appendix
\section{Photon emission rate}\label{App:photon_emission_rate}

The photon emission rate into the radiation mode $\mathrm{R}$ at time $t$ is determined by 
\begin{align}
\label{eq:P(t)}
    P(t)&=\frac{d}{dt}\textrm{Tr}\left\{\hat{{N}}_R \hat{\rho}_T(t)\right\}
\end{align}
where the number operator $\hat{{N}}_R=\hat{a}^{\dagger}_R\hat{a}_R$. Using the linearity of the trace, we move the time derivative inside and use the equation of motion of the density operator to obtain
\begin{eqnarray}
 P(t)&=&\ \ \textrm{Tr}\left\{\hat{{N}}_R \frac{d}{dt}\hat{\rho}_\mathrm{T}(t)\right\}\nonumber\\
    &=&
    -\frac{i}{\hbar}\textrm{Tr}\left\{\hat{{N}}_R\, [\hat{H}_\mathrm{0},\hat{\rho}_\mathrm{T}(t)]+\hat{{N}}_R\, [\hat{H}_\mathrm{L}(t),\hat{\rho}_\mathrm{T}(t)] \right\}\nonumber\\
    &=&
    -\frac{i}{\hbar}\textrm{Tr}\left\{\hat{{N}}_R\, [\hat{H}_\mathrm{L}(t),\hat{\rho}_\mathrm{T}(t)] \right\}.\label{Eq:App_PER}
\end{eqnarray}
Here we use the cyclic property of the trace which leads to $\hat{{N}}_R[\hat{H}_\mathrm{0},\hat{\rho}_\mathrm{T}(t)]=0$ since $\hat{{N}}_R$ commutes with $\hat{H}_\mathrm{0}$. Initially the radiation mode is not occupied.
The number operator $\hat{N}_R$ conserves the number of excitations and $\hat{H}_\mathrm{L}$ contains terms with at most one of the operators $\hat{a}_R$ or $\hat{a}_R^\dagger$.
Therefore, in perturbation theory (and applying the rotating wave approximation), the dominant contribution comes from terms that contain four interactions $\hat{H}_\mathrm{L}$ (two with the radiation field and two with the classical field). 

To evaluate the photon emission rate, we insert the third-order expression in the perturbative expansion of the density operator with respect to $\hat{H}_\mathrm{L}(t)$ in Eq.~\eqref{Eq:App_PER}. 
After rearranging the commutators inside the trace we obtain
\begin{equation}
\label{Eq:Pt_Alex}
\begin{split}
P(t)&= \Big(\frac{-\mathrm{i}}{\hbar}\Big)^4
\int_{t_0}^t\!\! d\tau_3
\int_{t_0}^{\tau_3}\!\!\!\! d\tau_2
\int_{t_0}^{\tau_2}\!\!\!\! d\tau_1
\\
&\mathrm{Tr}\Big\{
[[[[\hat{N}_R,\hat{H}'_\mathrm{L}(t)],\hat{H}'_\mathrm{L}(\tau_3)],\hat{H}'_\mathrm{L}(\tau_2)],\hat{H}'_\mathrm{L}(\tau_1)]
\hat{\rho}(0)
\Big\}
\end{split}
\end{equation}
where the prime denotes an operator in the interaction representation with respect to  $\hat{H}_0=\hat{H}_\mathrm{M}+\hat{H}_\mathrm{F}$.   
Explicitly, we have 
\begin{eqnarray}
\label{eq:AppP(t)tauIntegrals}
    \hat{H}^{'}_\mathrm{L}(t)&=&\hat{U}_0^{\dagger}(t)\hat{H}_\mathrm{L}(t)\hat{U}_0(t)\nonumber\\
    &=&-\hat{\bfmu}'(t)\cdot\mathbf{E}_{I}(t) -\hat{\bfmu}'(t)\cdot\hat{\mathbf{E}}'_{R}(t)\nonumber\label{Eq.Perturbation_interaction_pic}
\end{eqnarray}
where 
\begin{equation}
\hat{\bfmu}'(t)=\hat{U}_0^{\dagger}(t)\hat{\bfmu}\hat{U}_0(t)= \exp(\mathrm{i} \hat{H}_\mathrm{M}t/\hbar)\hat{\bfmu}\exp(-\mathrm{i} \hat{H}_\mathrm{M}t/\hbar)
\end{equation}
and 
\begin{equation}
  \hat{\mathbf{E}}'_R(t)=E_R\hat{a}_Re^{-i\omega_Rt}\bfepsilon_R+E_R^* \hat{a}_R^{\dagger}e^{i\omega_Rt}\bfepsilon_R 
\end{equation}
where we have used that $\hat{U}_0^{\dagger}(t)\hat{a}_R\hat{U}_0(t)=\hat{a}_Re^{-i\omega_Rt}$ and $\hat{U}_0^{\dagger}(t)\hat{a}_R^{\dagger}\hat{U}_0(t)=\hat{a}_R^{\dagger}e^{i\omega_Rt}$.

Of the 32 terms in Eq.~(\ref{eq:AppP(t)tauIntegrals}), only three, along with their complex conjugates, contribute significantly to the spontaneous light emission process, while the remaining terms can be neglected.  We can identify these 6 terms  based on the following arguments (see also the discussion before Eq.~(9.10) in Ref.~\onlinecite{mukamel1995principles}): (a) two interactions should be with the classical field and two with the quantum mode $R$, (b) each bra and ket state should have one interaction with the classical field and one with the quantum mode, (c) since initially both the quantum mode and the system are in their ground state, within the rotating wave approximation, the first interaction should be an interaction (system excitation) with the classical field, and the last interaction should be an  with the quantum mode (deexcitation of the system). 
Introducing the usual time intervals $t_3=t-\tau_3$, $t_2=\tau_3-\tau_2$ and $t_1=\tau_2=\tau_1$ we find that
Eq.~\eqref{Eq:Pt_Alex} is approximately given by Eq.~\eqref{Eq:Approx_Pt} of the main manuscript. 

\section{Dyadic HOPS}\label{AppI}
In the dyadic HOPS formalism, the third order response function $(R(t_2,t_3))$ in Eq.~\eqref{Eq:Generic_Dyadic_R} is constructed from the time-evolved third order density matrix using individual bra and ket contributions,\cite{mukamel1995principles,Chen2022JCP}
\begin{equation}
    \hat{\rho}^{(3)}(t)=\ket{\Phi_\mathrm{K}(t)}\bra{\Phi_\mathrm{B}(t)}
\end{equation}
where $\ket{\Phi_K(t)}$ and $\ket{\Phi_B(t)}$ represent the time-evolution of the ket and bra states, respectively, in the combined system-bath Hilbert space. In the zero-temperature limit the initial material density matrix is
\begin{equation}
\hat{\rho}(0) =\hat{\rho}_\mathrm{S}^{\textrm{eq}}\otimes\hat{\rho}_\mathrm{B}^{\textrm{eq}} = \ket{g}\ket{\mathbf{0}}\bra{\mathbf{0}}\bra{g}
\end{equation}
where the thermal environment is in the ground state ($\ket{\mathbf{0}}$).The corresponding time-evolution equations for bra and ket states
are given by
\begin{eqnarray}
    \ket{\Phi_\mathrm{K}(t)}&=&\hat{U}_\mathrm{M}(t_2+t_3)\hat{V}^{+}_{I}\ket{g}\ket{\mathbf{0}}\label{Eq:ket_Phi}\\
    \ket{\Phi_\mathrm{B}(t)}&=&\hat{U}_\mathrm{M}(t_3) \hat{V}^-_R\hat{U}_\mathrm{M}(t_2)\hat{V}^{+}_{I}\ket{g}\ket{\mathbf{0}}\label{Eq:bra_Phi}
\end{eqnarray}
  and the response function $R(t)$ can be rewritten as
\begin{align}
   R(t_2,t_3)&= \textrm{Tr}\{\hat{F}\ket{\Phi_\mathrm{K}(t)}\bra{\Phi_\mathrm{B}(t)}\}\\
   &=\bra{\Phi_\mathrm{B}(t)}\hat{F}\ket{\Phi_\mathrm{K}(t)}\label{Eq:response_func_orig}.
\end{align}

We evaluate the dyadic response function using the non-Markovian quantum state diffusion (NMQSD) formalism.\cite{Diosi1997} The first step is to introduce an identity operation on each side of $\hat{F}$ in the form of an integral over a complete set of Bargmann coherent states for the bath
\begin{align}
    \hat{\mathbb{I}} = &\int  dM(\mathbf{z})  \ket{\mathbf{z}}\bra{\mathbf{z}} \\ 
    dM(\mathbf{z}) &= \prod_{n\lambda}d^2z_{n,\lambda}\frac{e^{-|z_{n,\lambda}|^2}}{\pi}
\end{align}
and then use the reproducing property of coherent states\cite{robert2021coherent} to remove one of the integrals, giving
\begin{equation}
   R(t_2,t_3)=\int dM(\mathbf{z})\bra{\phi_\mathrm{B}(t;\mathbf{z^*})}\hat{F}\ket{\phi_\mathrm{K}(t;\mathbf{z^*})}
    \label{Eq:NMQSD_response}
\end{equation}
where $\ket{\phi_\mathrm{B/K}(t;\mathbf{z^*})}=\braket{\mathbf{z^*}|\Phi_{B/K}(t)}$. Following Ref.~\onlinecite{Diosi1997}, the integral in Eq.~\eqref{Eq:NMQSD_response} can be equivalently described by an ensemble average $(\text{denoted by} \mathcal{M}_{\mathbf{z}}[\cdot])$ over a stochastic process $\mathbf{z^*}$,
\begin{equation}
   R(t_2,t_3)=\mathcal{M}_{\mathbf{z}}[\bra{\phi_\mathrm{B}(t;\mathbf{z^*})}\hat{F}\ket{\phi_\mathrm{K}(t;\mathbf{z^*})}]\label{Eq:response_hilbert}
\end{equation}
  where elements in $\mathbf{z^*}$ are denoted by $z_{n,t}$ and the stochastic process is defined by $\mathcal{M}[z_{n,t}]=0$,  $\mathcal{M}[z_{n,t} z_{n,s}]=0$, and $\mathcal{M}[z^*_{n,t} z_{n,s}]=\alpha_n(t-s)$.
Here, we have provided the dyadic NMQSD derivation at zero-temperature, however, as has been discussed previously,\cite{Chen2022JCP} the finite-temperature formulation can be derived using the thermofield methods.\cite{Strunz1998PRA, Gerhard2015JCP, Semenoff1983} The result is a system of equations that, for Hermitian $\hat{L}$-operators, is equivalent to the preceding zero-temperature version where temperature dependence appears only in the bath correlation function given by Eq.~\eqref{eq:bath_corr}.\\
We can write the response function (Eq.~\eqref{Eq:response_hilbert}) in dyadic framework by defining a wavefunction in a doubled system Hilbert space,
\begin{equation}
    \ket{\tilde{\psi}(t;\mathbf{z}^*)}=\begin{pmatrix}
        \ket{\phi_{\mathrm{K}}(t;\mathbf{z}^*)}\\
        \ket{\phi_{\mathrm{B}}(t;\mathbf{z}^*)}\label{Eq:dyadic_wf}
    \end{pmatrix}
\end{equation}
resulting in,
\begin{equation}
    R(t)=\int dM(\mathbf{z})\bra{\tilde{\psi}(t;\mathbf{z}^*)}\tilde{F}\ket{\tilde{\psi}(t;\mathbf{z}^*)}
\end{equation}
with, $\tilde{F}=\begin{pmatrix}
    0 & 0\\
    \hat{F} & 0
\end{pmatrix}$. Notice that in the dyadic formalism the response function becomes the expectation value of the (non-Hermitian) $\tilde{F}$ operator that connects the bra component ($\ket{\phi_{\mathrm{B}}(t;\mathbf{z}^*)}$) with the ket component ($\ket{\phi_{\mathrm{K}}(t;\mathbf{z}^*)}$).

\section{Time dependent fluorescence spectrum}\label{App:TDF}
\begin{figure}
\includegraphics[width=1\linewidth]{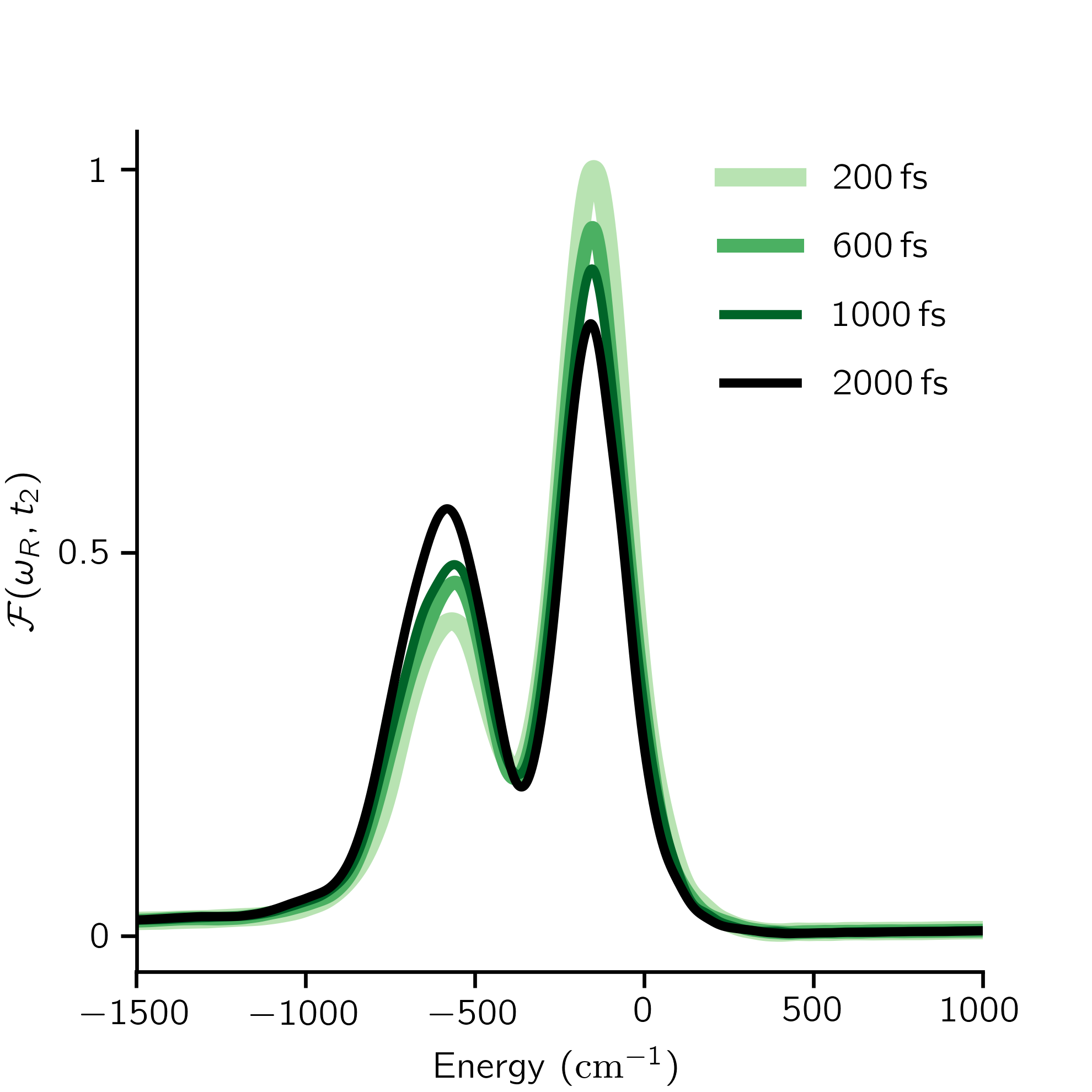}
\caption{Demonstration of dynamical Stokes shift using time dependent fluorescence spectrum. We simulate fluorescence with different values of $t_2\,(\textrm{fs})$ for 4-site chain where $E_n=0$ for three sites and for the forth site it is $-500\,\textrm{cm}^{-1}$. Parameters: nearest neighbor coupling $ J_{n,n\pm1}=-50 \textrm{ cm}^{-1}$, $\lambda = 50 \textrm{ cm}^{-1}$, $\gamma=50 \textrm{ cm}^{-1}$, $T=295 \textrm{ K}$, $t_3^\textrm{{max}}=500\,\textrm{fs}$, and $\omega_{\textrm{res}}=1.25\times10^{-4}\,\textrm{fs}^{-1}$. The convergence parameters are given in Table~\ref{tab:CS}.}\label{Fig:4site_defect}
\end{figure}

The current DadHOPS formalism is capable of capturing the time-evolution of the fluorescence spectrum. To illustrate this, we consider a 4-site linear chain with a defect: three sites have identical excitation energies, while the fourth site is detuned by $500\,\textrm{cm}^{-1}$ to a lower energy. As the waiting time $t_2$ increases, we expect population relaxation into the lower-energy defect site, leading to a corresponding increase in the fluorescence signal at that energy. This behavior is clearly observed in Fig.~\ref{Fig:4site_defect}. As $t_2$ increases from $200\,\textrm{fs}$ (light green) to $2000\,\textrm{fs}$ (dark green), the fluorescence peak near 0 energy decreases, while the peak near $-500\,\textrm{cm}^{-1}$ grows in intensity. For visual clarity, all spectra are scaled relative to the spectrum at $t_2 = 200\,\textrm{fs}$.

\section{Size-invariant absolute spectrum}\label{App:Abs_spectrum}
\begin{figure}
\includegraphics[width=1\linewidth]{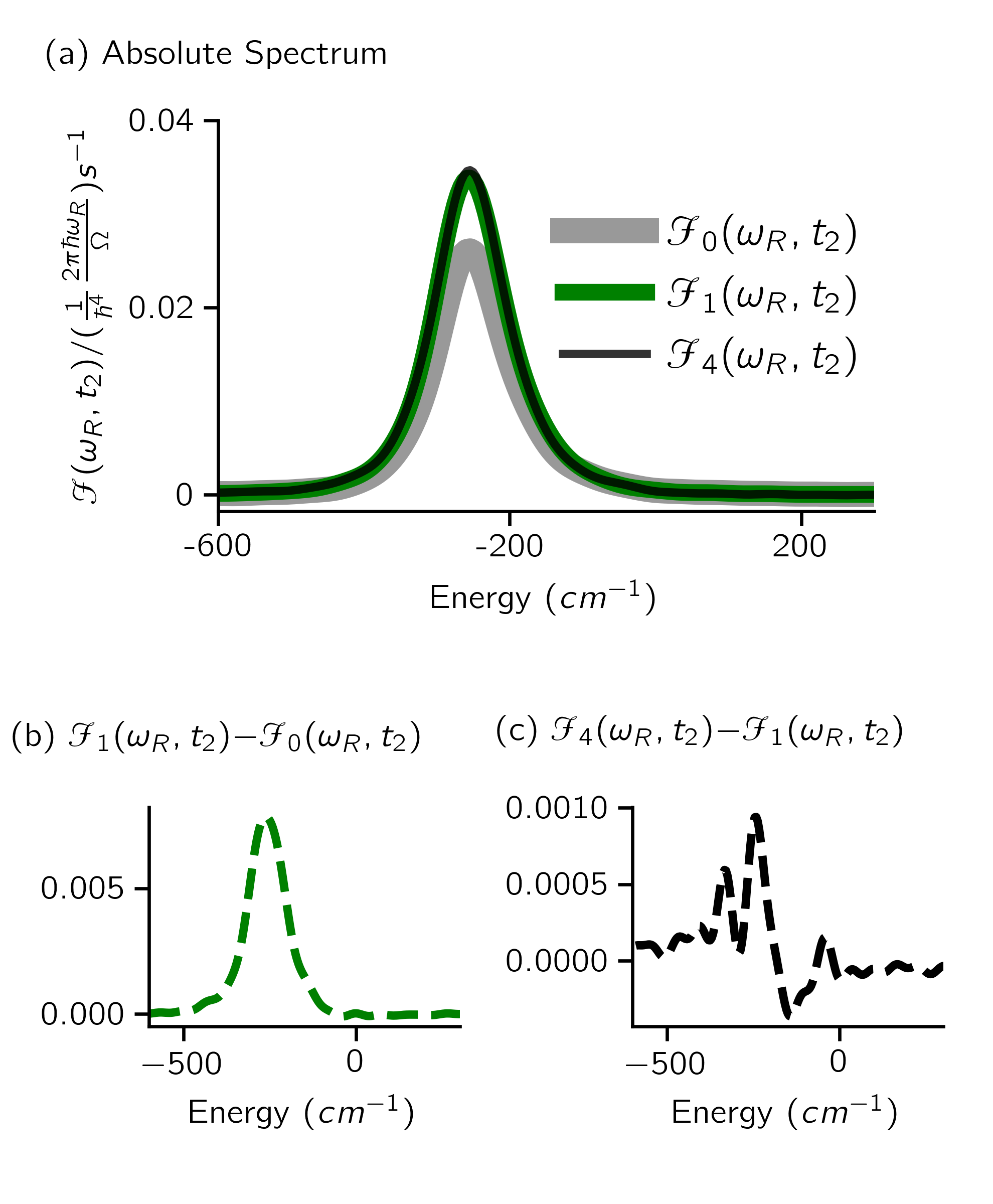}
\caption{Population contributions ($r_{\{i \vert i\}}$) alone are insufficient for the absolute fluorescence spectrum, but adding nearest neighbors ($r_{\{i \vert i \pm 1\}}$) is sufficient. (a) Absolute spectrum with $\mathdutchcal{F}$$_0(\omega_R, t_2)$ (grey), $\mathdutchcal{F}$$_1(\omega_R, t_2)$ (green), and $\mathdutchcal{F}$$_4(\omega_R, t_2)$ (black). (b) Difference between $\mathdutchcal{F}$$_0(\omega_R, t_2)$ (grey) and $\mathdutchcal{F}$$_1(\omega_R, t_2)$. (c) Difference between $\mathdutchcal{F}$$_1(\omega_R, t_2)$ (grey) and $\mathdutchcal{F}$$_4(\omega_R, t_2)$.   Parameters: $ J_{n,n\pm1}=-100 \textrm{ cm}^{-1}$, $\lambda = 35 \textrm{ cm}^{-1}$, $\gamma=50 \textrm{ cm}^{-1}$, $T=295 \textrm{ K}$, $t_2=400\,\textrm{fs}$, $t_3^\textrm{{max}}=500\,\textrm{fs}$, and $\omega_{\textrm{res}}=1.25\times10^{-4}\,\textrm{fs}^{-1}$. The convergence parameters are given in Table~\ref{tab:CS}.}\label{Fig:AppC}
\end{figure}

The absolute (rather than normalized) fluorescence spectrum can also be efficiently calculated with a local decomposition.
In Fig.~\ref{Fig:AppC} we present the absolute spectrum corresponding to the results in Sec.~\ref{Sec:Eff_MC_sampling}. Unlike the relative spectrum, the absolute spectrum $\mathdutchcal{F}$$_0(\omega_R, t_2)$ (grey line, Fig.~\ref{Fig:AppC}a) has a reduced peak height compared to more exact calculations. Expanding the window of bra excitations to include sites neighboring the ket site ($\mathdutchcal{F}$$_1(\omega_R, t_2)$) enhances the peak magnitude (green line, Fig.~\ref{Fig:AppC}a). However, further expansion of the bra excitation window around the ket site provides no further increase in peak magnitude as seen in $\mathdutchcal{F}$$_4(\omega_R, t_2)$ (black line, Fig.~\ref{Fig:AppC}a). Figs.~\ref{Fig:AppC}b and c further quantify the impact of changing the window of bra-side excitations on the peak height. 

\section{Generalized algorithm for an arbitrary aggregate maintaining size-invariance}\label{AppA}

For an arbitrary system Hamiltonian where the most important bra excitation operators cannot be identified by eye, we have developed an ansatz to identify the most important bra operators while ensuring size-invariant scaling:

\section*{Algorithm}

The total excitation operator for the ket state can be decomposed into a sum over a user-selected, arbitrary set of local excitation operators cf.~Eq.\eqref{Eq:EOD_first_step},
\begin{align}
\label{eq:appD_ketEOD}
    V^{+}_I = \sum_{\mathbf{d}_{\textrm{K}}\in \mathcal{D}_{\textrm{K}}} \hat{\sigma}^+_{\mathbf{d}_{\textrm{K}}}
\quad
\mathrm{where}
\quad
    \hat{\sigma}^+_{\mathbf{d}_{\textrm{K}}} = \sum_{n \in\mathbf{d}_{\textrm{K}} } \ \frac{\bfmu_n \cdot\bfepsilon_{I}}{D_I} \,\vert n\rangle\langle g \vert. 
\end{align}
In Section \ref{Sec:Eff_MC_sampling}, the set of local excitation operators are chosen to be the single-site excitation operators. Here, we will generalize to allow for any set of local excitation operators satisfying Eq.~\eqref{eq:appD_ketEOD}. For computational efficiency, we recommend the user select a cluster size no longer than the expected delocalization extent.   

The Monte Carlo algorithm described in Section \ref{Sec:Eff_MC_sampling}, requires randomly selecting an initial ket excitation operator and then selecting $l_\mathrm{b}$ sites to define the bra excitation operator
\begin{equation}
    \hat{\sigma}^+_{\mathbf{d}'_{\textrm{B}}} = \sum_{n \in\mathbf{d}'_{\textrm{B}} } \ \frac{\bfmu_n \cdot\bfepsilon_{I}}{D_I} \,\vert n\rangle\langle g \vert.
\end{equation}
Below, we outline an algorithm that uses linear absorption calculations to efficiently sort the possible sites for the bra excitation operator by their expected contributions to the fluorescence calculation.

\renewcommand{\thesubsubsection}{\arabic{subsubsection}}
\subsubsection{Random Sampling of Ket State} 
    
Select a ket excitation operator $(\hat{\sigma}^+_{\mathbf{d}_{\textrm{K}}})$ for initial excitation of the cluster $ \mathbf{d}_\mathrm{K} \in \mathcal{D}_\mathrm{K}$ using unbiased random sampling.  

\subsubsection{Linear Absorption Simulation} 

For the selected ket excitation operator $(\hat{\sigma}^+_{\mathbf{d}_{\textrm{K}}})$, run approximately 100 linear absorption trajectories, each for a total time of \( T_\mathrm{\max} \), and calculate the time-dependent correlation function:  
\begin{equation}
C_{\mathbf{d}_\mathrm{K}}(t;\mathbf{z^*})=I\bra{\tilde{\psi}_{\mathbf{d}_\mathrm{K}}(t;\mathbf{z^*})}\tilde{F}\ket{\tilde{\psi}_{\mathbf{d}_\mathrm{K}}(t;\mathbf{z^*})}
\end{equation}
where the wave function
\begin{equation}
\ket{\tilde{\psi}_{\mathbf{d}_\mathrm{K}}(t;\mathbf{z^*})}=\tilde{G}(t;\mathbf{z^*}) \ket{\tilde{\psi}_{\mathbf{d}_\mathrm{K}}(0)}\label{Eq:App_WF}
\end{equation}
is propagated using the initial state  
\begin{equation}
\ket{\tilde{\psi}_{\mathbf{d}_\mathrm{K}}(0)} = \frac{1}{\sqrt{2}}\begin{pmatrix}\hat{\sigma}^+_{\mathbf{d}_\textrm{K}}\ket{g}\\\ket{g} \end{pmatrix},  
\end{equation}
and normalization factor 
\begin{equation}
I=||\ket{\tilde{\psi}_{\mathbf{d}_K} (0)}||^2\cdot||\ket{\tilde{\phi}(0)}||^2   
\end{equation}
where 
\begin{equation}
\ket{\tilde{\phi}(0)} = \begin{pmatrix}\ket{g}\\\ket{g} \end{pmatrix}.  
\end{equation}
Write the states as  \begin{equation}
        \ket{\tilde{\psi}^{(\vec{0})}(t;\mathbf{z^*})}= \sum_{n=1}^{2N} c_n(t;\mathbf{z^*}) \ket{n}\label{Eq:App_WF_coeff}
        \end{equation}
        and store the coefficient $c_n(t)$ for $t=0,\ \Delta t,\ 2\Delta t,\dots , T_\mathrm{max}$.
Note: Since we never excite the bra state for absorption, half of the coefficients in Eq.~\eqref{Eq:App_WF_coeff} will be zeros. 
For each pigment $n$, compute $P_n=\mathcal{M}_{\mathbf{z}}[P_n(\mathbf{z^*})]$ with 
        \begin{equation}
    P_n(\mathbf{z^*})=\big<|c_n(t;\mathbf{z^*})|^2 C_{\mathbf{d}_\mathrm{K}}(t;\mathbf{z^*})\big>_t/\big<C_{\mathbf{d}_\mathrm{K}}(t;\mathbf{z^*}) \big>_t
    \end{equation}
     where $\big<\cdots \big>_t$denoting average over time.

    \subsubsection{Selection of Bra Clusters} 
    
    Identify the \( l_b \) sites with the highest values of \( P_n \) to define the bra cluster \( \mathbf{d}'_\textrm{B} \).  

    \subsubsection{Fluorescence Calculation}  
    
    Simulate fluorescence using the selected ket cluster \( \mathbf{d}_\mathrm{K} \) and bra cluster \( \mathbf{d}'_\mathrm{B} \).  Note: Although not required, we recommend keeping \(l_b \) consistent for every bra cluster. 

    \subsubsection{Repeat}  
    To create an ensemble for response functions, go back to step 1 and perform step 2 only when a new cluster in step 1 is selected.  

\subsection*{Convergence Check}
Since the bra cluster size \( l_b \) is a tunable parameter, ensure convergence by:  
\begin{itemize}
    \item Iteratively including additional sites beyond \( l_b \) based on \( P_n \).  
    \item Repeating Steps 1, 3, 4, and 5 while skipping step 2.  
    \item Comparing spectra between iterations and computing the error.
\end{itemize}
If the error falls within an acceptable range, the spectrum is considered converged. Otherwise, continue expanding \( \mathbf{d}'_\mathrm{B} \) until convergence is achieved.

\section{Fluorescence spectrum for 1D H-aggregates}\label{App:H_agg}

\begin{figure}
\includegraphics[width=1\linewidth]{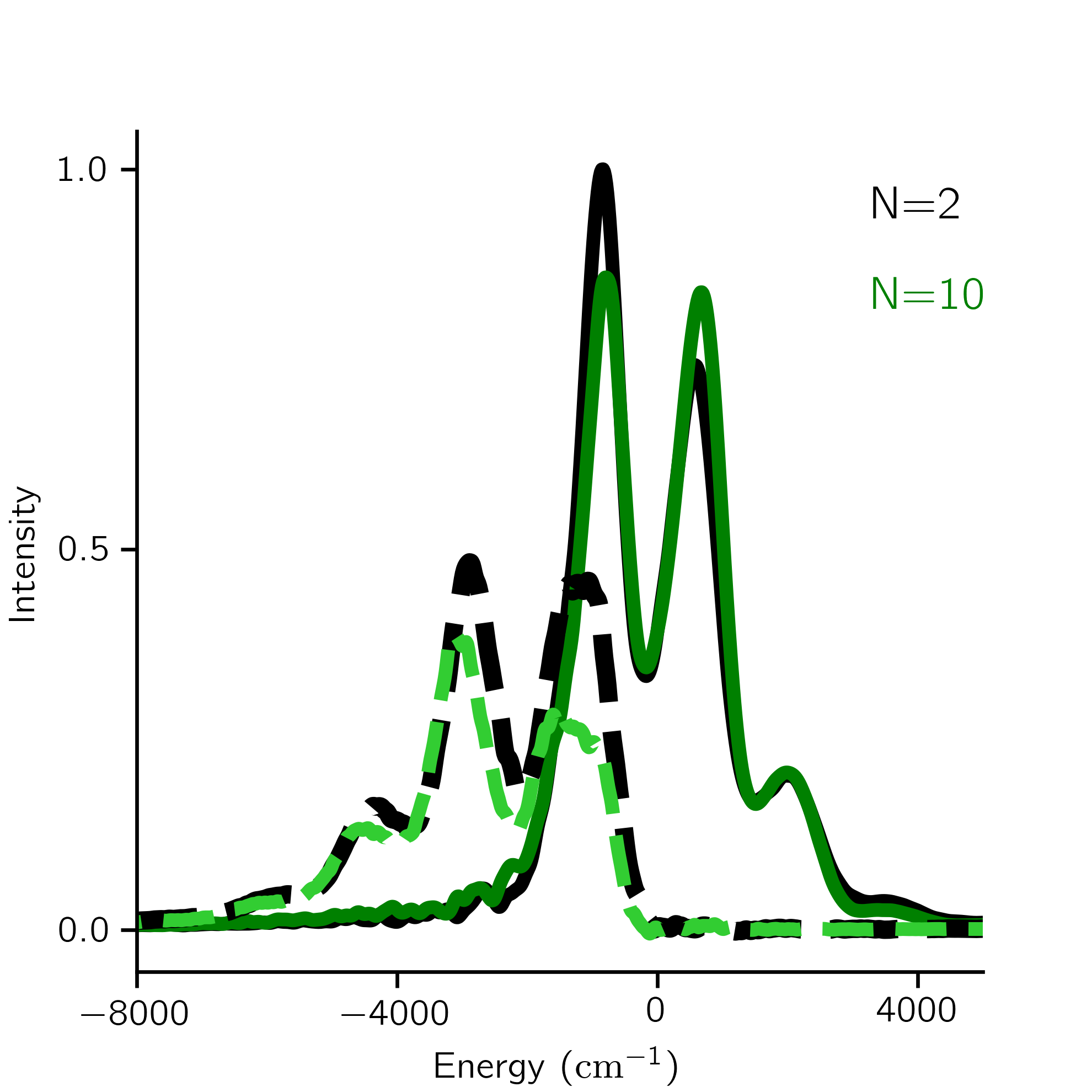}
\caption{Exact methods like HOPS are required to reproduced the correct lineshape. Fluorescence (dashed line) and Absorption (solid line) spectrum for H-aggregates with chain length $N=2$ (black) and $N=10$ (green).  Parameters: $T=298 \textrm{K}$, $t_2=400\,\textrm{fs}$, $t^{\textrm{max}}_3=500\,\textrm{fs}$, $g_1=(2.05\times10^4-2.5\times10^3i)\, \textrm{cm}^{-2}$, $\gamma_1=50\, \textrm{cm}^{-1}$, $g_2 = 1.6\times10^6\, \textrm{cm}^{-2}$, $\gamma_2 = (100 + 1550i)\, \textrm{cm}^{-1}$,  $\omega_{\textrm{res}}=5\times10^{-4}\,\textrm{fs}^{-1}$, and nearest-neighbor coupling $ J_{n,n\pm1}=300\, \textrm{cm}^{-1}.$ 
The convergence parameters are given in Table~\ref{tab:CS}.}\label{Fig:Hagg}
\end{figure}

The entanglement between electronic and vibrational degrees of freedom introduces additional dipole-allowed transitions that are absent in a purely electronic picture. This effect can render the 0-0 peak optically bright in systems that would otherwise exhibit a dark state, such as H-aggregates. Figure~\ref{Fig:Hagg} shows the normalized (with respect to the dimer absorption peak height) fluorescence spectra for H-aggregates with chain lengths $N=2$ (black dashed line) and $N=10$ (green dashed line). In both cases, the 0-0 peak is present but less intense than the 0-1 peak. However, for $N=10$, the relative suppression of the 0-0 peak is more pronounced, reflecting the system's greater delocalization extent. For reference we also plot the absorption spectrum for chain lengths $N=2$ (black solid line) and $N=10$ (green solid line).

\section{Convergence parameters}
In Table.~\ref{tab:CS}, we list all the convergence parameters and their corresponding values used to calculate the response functions in the figures of this manuscript.
\begin{table*}[t]
    \centering
    \caption{Convergence parameters for calculations performed for each figure.}
    \begin{tabular}{|c|c| c| c |c |c |c |c|}
        \hline
        Figure & $k_{\textrm{max}}$ & $\Delta t$ fs & $\delta_A$ & $\delta_S$ & $u_S$ & $f_{dis}$ &$N_{ens}$  \\
        \hline
        3 (black line) & 8 & 0.4 & -- & -- & -- & -- & 10000
        \\
        \hline
        3 (light blue and lime green lines) & 10 & 0.4 & -- & -- & -- & -- & 20000
        \\
        \hline
        3 (dark blue and dark green lines) & 10 & 0.4 & -- & -- & -- & -- & 50000
        \\
        \hline
        4  & 15 & 0.25 & -- & -- & -- & -- & 100000 \\
        \hline
        5  & 15 & 0.25 & 0.0001 & 0.0001 & 2 & 0.01 & 30000 \\
        \hline
        6  & 15 & 0.25 & 0.0001 & 0.0001 & 2 & 0.01 & 2000 \\
        \hline
        7  & 15 & 0.25 & 0.0001 & 0.0001 & 2 & 0.01 & 2000 \\
        \hline
        8  & 10 & 0.10 & 0.01 & 0.05 & 4 & 0.01 & 10000 \\
        \hline
        9  & 12 & 1.00 & - & - & - & - & 10000 \\
        \hline
        10  & 15 & 0.25 & 0.0001 & 0.0001 & 2 & 0.01 & 30000 \\
        \hline
        11  & 10 & 0.10 & 0.01 & 0.05 & 4 & 0.01 & 20000 \\
        \hline
    \end{tabular}
    \label{tab:CS}
\end{table*}

\section*{References}
	\bibliography{fluorescence}
	\bibliographystyle{apsrev}
\end{document}